\documentclass[12pt]{article}
\usepackage{amsfonts,amssymb,bm,amsmath,cite}
\usepackage{array, booktabs} 
\usepackage[dvipdfmx]{graphicx}
\usepackage{physics} 
\numberwithin{equation}{section}
\usepackage{xcolor}

\usepackage{hyperref} 
\hypersetup{pdftitle={Spacetime Emergence in IIB Matrix Model}, colorlinks=true,linkcolor=blue,filecolor=magenta,citecolor = blue,anchorcolor = blue, urlcolor=blue, bookmarks=true}
\usepackage[colorinlistoftodos,shadow]{todonotes}

\renewcommand{\vev}     [1]{\ensuremath{\langle #1 \rangle}}
\newcommand  {\rf}      [1]{(\ref{#1})}

\newcommand  {\mf}         {\ensuremath{m_{\mbox{\scriptsize f}}}}

\newcommand{\nn}{\nonumber}
\newcommand{\bbR}{{\mathbb R}}
\newcommand{\bbC}{{\mathbb C}}
\newcommand{\bbZ}{{\mathbb Z}}

\begin{document}
\setlength{\oddsidemargin}{0cm}

\begin{titlepage}
\renewcommand{\thefootnote}{\fnsymbol{footnote}}
\begin{normalsize}
\begin{flushright}
\begin{tabular}{r}
April 2026 \\
KEK-TH-2826
\end{tabular}
\end{flushright}
  \end{normalsize}

~~\\

\vspace{1cm}

\begin{Large}
       \begin{center}
         {The emergence of (3+1)-dimensional expanding spacetime\\
           from complex Langevin simulations of the Lorentzian type IIB matrix model
         with deformations}
       \end{center}
    \end{Large}
\vspace{0.7cm}

\begin{center}
           Konstantinos~N. A{\sc nagnostopoulos}$^{1)}$\footnote
            {
e-mail address : 
konstant@mail.ntua.gr},
           Takehiro~A{\sc zuma}$^{2)}$\footnote
            {
e-mail address : 
azuma@mpg.setsunan.ac.jp},
           Mitsuaki~H{\sc irasawa}$^{3,4)}$\footnote
           {
e-mail address : mitsuaki.hirasawa@unimib.it},
		  Jun~N{\sc ishimura}$^{5,6)}$\footnote
		              {
		  e-mail address : 
  jnishi@post.kek.jp}, 
             Stratos~P{\sc apadoudis}$^{1)}$\footnote
              {
  e-mail address : 
  sp10018@central.ntua.gr},
           and
           Asato~T{\sc suchiya}$^{7)}$\footnote
            {
e-mail address : 
tsuchiya.asato@shizuoka.ac.jp}\\

\vspace{2cm}

$^{1)}$ {\it Physics Department, School of Applied Mathematical and Physical Sciences, }\\
              {\it  National Technical University of Athens, Zografou Campus, GR-15780 Athens, Greece}\\

$^{2)}$ {\it Institute for Fundamental Sciences, Setsunan University,}\\
              {\it 17-8 Ikeda Nakamachi, Neyagawa, Osaka, 572-8508, Japan}\\

$^{3)}$ {\it Department of Physics, Universit\`{a} degli Studi di Milano-Bicocca,}\\
              {\it Piazza della Scienza 3, I-20126 Milano, Italy}\\
              
$^{4)}$ {\it Sezione di Milano-Bicocca, Istituto Nazionale di Fisica Nucleare (INFN),}\\
              {\it Piazza della Scienza 3, I-20126 Milano, Italy}\\

$^{5)}$ {\it KEK Theory Center, Institute of Particle and Nuclear Studies,}\\
              {\it High Energy Accelerator Research Organization,}\\
              {\it 1-1 Oho, Tsukuba, Ibaraki 305-0801, Japan}\\
               
$^{6)}$ {\it Graduate Institute for Advanced Studies, SOKENDAI,}\\
              {\it 1-1 Oho, Tsukuba, Ibaraki 305-0801, Japan}\\
                 
$^{7)}$ {\it Department of Physics, Shizuoka University,} \\
              {\it 836 Ohya, Suruga-ku, Shizuoka 422-8529, Japan}\\
\end{center}

\hspace{3cm}

\clearpage
\begin{abstract}
\noindent
The Lorentzian type IIB matrix model is a promising candidate for a nonperturbative formulation
of superstring theory. In this model,
the eigenvalue distribution of the $N\times N$ bosonic matrices $A_\mu$ $(\mu = 0 , \ldots , 9)$
represents an emergent spacetime,
which is determined by the dynamics of the model
in the large-$N$ limit.
Here we perform numerical simulations of the model
overcoming the sign problem by the complex Langevin method with the matrix size $N$
up to $128$.
In order to avoid the singular drift problem due to the Pfaffian, which
appears after integrating out the fermionic matrices,
we deform the model in a manner inspired by the supersymmetric deformation,
which is used to define the ``polarized type IIB matrix model'' in the Euclidean case.
We find that the deformed model exhibits
a phase in which (3+1)-dimensional expanding spacetime emerges
with both space and time being smooth and real.
\end{abstract}
\vfill
\end{titlepage}
\vfil\eject

\tableofcontents

\setcounter{footnote}{0}

\section{Introduction}
\label{sec: intro}
The type IIB matrix model \cite{Ishibashi:1996xs} (or the IKKT matrix model)
is a promising candidate for a nonperturbative formulation of superstring theory.
In this model, spacetime\cite{Aoki:1998vn,Aoki:1998bq}---as well as
gauge fields \cite{Iso:1999xs} and
matter fields \cite{Aoki:2014cya,Hatakeyama:2019jyw} on it---
is expected to emerge dynamically in the large-$N$ limit.
Making the matrix size $N$ finite
serves as a regularization of the theory analogous to the lattice regularization
in gauge theories. 

As in lattice gauge theories,
Monte Carlo simulations
have been used to investigate the dynamical properties of this model.
One promising scenario explored up to now
is the emergence
of a phenomenologically viable cosmology, featuring a (3+1)-dimensional expanding spacetime.
In this scenario, the extra dimensions remain small due to the dynamics of the model,
eliminating the need for additional parameters typically required in conventional compactification
scenarios.
This raises the possibility that our universe could emerge from the model without encountering
the string landscape problem.

The realization of such a scenario was first discussed in Ref.~\cite{Aoki:1998vn}
within the context of the Euclidean model. In this model, classical solutions are
diagonal up to SU($N$) symmetry, and the diagonal elements represent spacetime coordinates.
By perturbing around these configurations, it was argued that fermionic zero modes
play a crucial role in the dimensional reduction of the emergent spacetime.
A nonperturbative study of the Euclidean model \cite{Nishimura:2011xy}
based on the Gaussian expansion method
demonstrated the spontaneous breaking of the SO($10$) rotational symmetry to SO($3$).
Monte Carlo simulations confirmed this
finding \cite{Anagnostopoulos:2015gua,Anagnostopoulos:2020xai} and
highlighted the importance of fermionic dynamics. Integrating out the fermions yields
a Pfaffian with a highly oscillatory phase. Its fluctuations are speculated to favor
lower-dimensional spaces \cite{hep-th/0003223,Nishimura:2000wf}.
In the absence of such fluctuations, the rotational symmetry remains
unbroken \cite{hep-th/9811220,Ambjorn:2000dx,Ambjorn:2000bf}.
Studies of toy models, employing either the Gaussian expansion
method \cite{Nishimura:2004ts,Aoyama:2010ry} or numerical
simulations \cite{Anagnostopoulos:2001yb,Anagnostopoulos:2010ux,Anagnostopoulos:2011cn,Anagnostopoulos:2013xga,Anagnostopoulos:2017gos}, have provided further support for this scenario.
For a comprehensive review, see Ref.~\cite{Anagnostopoulos:2022dak}. 

The Lorentzian model was not investigated until much later due to the severe sign problem
arising from the factor $e^{iS}$ in the path integral.
The first paper on the Lorentzian model \cite{Kim:2011cr}
and subsequent papers \cite{Ito:2013qga,Ito:2013ywa,Ito:2015mxa,Ito:2015mem,Ito:2017rcr}
suggested that dominant configurations
represent
a (3+1)-dimensional expanding spacetime.
However, Ref.~\cite{Aoki:2019tby} revealed that these configurations
possess a singular structure due to an artifact of the approximation employed to circumvent
the sign problem. This pointed to the necessity of simulating the model
without such approximations.
Indeed the complex Langevin method (CLM) \cite{Parisi:1983mgm,Klauder:1983sp} was used
to simulate the bosonic model (\emph{i.e.}, without fermions),
and the dominant configurations were found to become
smooth as one removes the approximation \cite{Nishimura:2019qal}.
Furthermore, it was found that
the Lorentzian model with extra quadratic terms \cite{Kim:2011ts,Kim:2012mw}
in the action
admits classical solutions
representing
smooth expanding configurations \cite{Hatakeyama:2019jyw}
although the dimensionality of space remains undetermined.
Similar deformations led also to other classical backgrounds,
where gravity can arise from the one-loop effective action at weak
coupling \cite{Steinacker:2021yxt,Steinacker:2023myp,Battista:2023glw,Kumar:2023bxg,Steinacker:2024unq,Manta:2024vol,Manta:2025inq,Manta:2025tcl}.
See also Refs.~\cite{Klinkhamer:2020xoi,Brahma:2021tkh,Brahma:2022dsd,Brahma:2022ikl,Klinkhamer:2022frp,Laliberte:2023bai,Brandenberger:2024ddi,Hattori:2024btt,Gohara:2025zfh,Ho:2025htr,Gass:2025bqr,Liao:2025yfb,Steinacker:2026qzk} for recent related works.

In this paper we attempt to investigate
the Lorentzian model with a Lorentz-invariant mass term
$S_\gamma = -\frac{N}{2} \gamma \Tr (A_\mu A^\mu)$, where $\gamma > 0$,
using the CLM.
It is well known that the CLM fails in some cases
due to the wrong convergence problem, which can be understood
from the arguments for justification \cite{Aarts:2009dg,Aarts:2009uq,Aarts:2011ax,Nishimura:2015pba,Nagata:2015uga,Nagata:2016vkn}.
In fact, this problem occurs in the present model due to singular drifts from the Pfaffian
since some eigenvalues of the associated matrix
come close to zero.
In order to avoid this problem,
we deform the model in a manner inspired by the supersymmetric (SUSY)
deformation \cite{Bonelli:2002mb} 
used to define the polarized
type IIB matrix model \cite{Hartnoll:2024csr,Komatsu:2024bop,Komatsu:2024ydh}
in the Euclidean case.\footnote{The polarized type IIB matrix model has
attracted much attention due to the discovery of the
holographic dual solutions in type IIB supergravity.
See also Refs.~\cite{Kumar:2022giw,Kumar:2023nya,Hartnoll:2025ecj,Chou:2025rwy}
for Monte Carlo studies.}
Within a certain parameter regime, we discover a phase in which
the space has small extent without SSB at early times and three out of nine
spatial directions start to expand at some point in time.
Furthermore, we find that both space and time turn out to be smooth and real.
Some preliminary results were
presented in Ref.~\cite{Anagnostopoulos:2022dak}
and proceedings articles \cite{Hatakeyama:2021ake,2201_13200,2205_04726,2212_10127,2407_03491}.
While the effects of the deformation have to be carefully examined in the next step,
our results strongly
suggest that the Lorentzian type IIB matrix model has remarkable
dynamical properties supporting the scenario for
the emergence of (3+1)-dimensional expanding spacetime in
nonperturbative string theory.

The rest of this paper is organized as follows.
In Section \ref{sec: IIBMM}, we briefly review the Lorentzian type IIB matrix model
and explain the motivations to add a Lorentz-invariant mass term.
In Section \ref{Sec_CLM}, we explain how one can apply the CLM to this model.
In Section \ref{results_sec}, we present our numerical results obtained by the CLM.
Section \ref{sec: conclusion} is devoted to a summary and discussions.
In Appendix \ref{sec:eliminating-effects-of-Lorentz=boosts}, we explain
how we remove the artifact of Lorentz boosts, which obscures our analyses of the
expanding spacetime.
In Appendix \ref{sec:SUSY-deformation},
we discuss the SUSY deformation of the type IIB matrix model
in the Lorentzian case
to clarify its relationship to the deformation used in the present work.

\section{Brief review of the Lorentzian type IIB matrix model}
\label{sec: IIBMM}

In this section, we briefly review the Lorentzian type IIB matrix model
and discuss, in particular, the equivalence to the Euclidean model,
which provides the motivation to add a Lorentz-invariant mass term.
We also discuss how one can extract the time evolution from
a matrix configuration.

\subsection{Equivalence to the Euclidean model}
\label{sec: def_IIBMM}

The action of the type IIB matrix model is given by \cite{Ishibashi:1996xs}
\begin{align}
\label{eq: action}
S&= S_\mathrm{b}+S_\mathrm{f} \ , \\
\label{eq: action_b}
S_\mathrm{b}&= - \frac{1}{4g^2} \Tr \left( [A_\mu,A_\nu][A^\mu,A^\nu]\right) \ ,\\ 
\label{eq: action_f}
S_\mathrm{f}&= - \frac{1}{2g^2} \Tr \left(\bar{\Psi}_{\alpha} (\Gamma^\mu)_{\alpha \beta}
[A_\mu,\Psi_{\beta}]\right) \ ,
\end{align}
where $A_\mu \ (\mu=0,\ldots,9)$ and $\Psi_\alpha$ ($\alpha=1,2,\dots,16$)
are $N\times N$ traceless Hermitian matrices,
which transform as a vector and a Majorana-Weyl spinor, respectively,
under SO($9,1$) transformations. 
The Lorentz indices are raised and lowered using
the Lorentzian metric
\begin{align}
\eta_{\mu \nu} = \textrm{diag} (-1,1,1,\dots,1)\ . \label{Lorentzian_metric}
\end{align}
$\Gamma^\mu$ are the ten-dimensional gamma matrices obtained after Weyl projection,
and $\bar\Psi = \Psi^T \mathcal{C}$, where $\mathcal{C}$ is the charge conjugation operator.
We set $g^2N=1$ without loss of generality.

The partition function is given by \cite{Kim:2011cr}
\begin{equation}
\label{eq: partition_func}
Z=\int dA d \Psi\,  e^{i(S_\mathrm{b}+ S_\mathrm{f})}
= \int dA\, e^{i S_\mathrm{b}}\, \mathrm{Pf} \mathcal{M}(A) \ ,
\end{equation}
where ${\cal M}$ is a $16 (N^2-1) \times 16(N^2-1)$ anti-symmetric matrix representing
the linear transformation 
\begin{align}
  \Psi_{\alpha} \to ({\cal M} \Psi)_{\alpha}
  =  (\mathcal{C}\Gamma^{\mu})_{\alpha \beta} [A_{\mu}, \Psi_{\beta}] \ ,
  \label{calM_def}
\end{align}
acting on the linear space of traceless complex $N \times N$ matrices $\Psi_{\alpha}$. 
One can easily prove that
the Pfaffian $\mathrm{Pf} \mathcal{M}(A)$ is real unlike the Euclidean case.

Since the integral \eqref{eq: partition_func} is not absolutely convergent,
here we define the model by analytic continuation as follows.\footnote{Note that
this procedure breaks Lorentz symmetry.
Recently Ref.~\cite{Asano:2024def} proposed
a new definition of the Lorentzian model, in which the Lorentz symmetry is
fixed by the Faddeev-Popov procedure.
In that case, the equivalence to the Euclidean model
does not hold as demonstrated numerically for $N=2$ in Ref.~\cite{2501_17798}.
It is considered, however, that the equivalence recovers in the $N \rightarrow \infty$
limit since the effect of the Faddeev-Popov determinant is suppressed at large $N$.}
Setting $\tilde{S}_\mathrm{b} = -i S_\mathrm{b}$ and rewriting the partition function
as $Z=\int dA e^{-\tilde{S}_\mathrm{b}}\mathrm{Pf} \mathcal{M}(A)$, we have
\begin{equation}
  \tilde{S}_\mathrm{b} =- \ \frac{i}{4}N\qty[-2\Tr({F}_{0i})^2 +\Tr({F}_{ij})^2]\ ,
  \label{LIKKT_boson}
\end{equation}
where $F_{\mu\nu}=i\comm{A_\mu}{A_\nu}$. The indices $i,j$ run over $1,2,\dots,9$.
We consider the contour deformation defined by \cite{Nishimura:2019qal}
\begin{align}
  A_0 = e^{ - i \frac{3\pi}{8} u} {\tilde A}_0, \ \ A_i = e^{ i  \frac{\pi}{8} u}
  {\tilde A}_i \ (i=1,2,\dots,9) \ , \quad
  0\leq u \leq 1,
  \label{wick_rotation_u}
\end{align}
where 
$\tilde A_\mu$ are Hermitian.
The action \eqref{LIKKT_boson} then becomes
\begin{align}
  \tilde{S}_\mathrm{b} =
  \frac{N}{4} \left[ 2 e^{i \frac{\pi}{2}  (1-u)} \textrm{Tr} ({\tilde F}_{0i})^2 +
    e^{- i \frac{\pi}{2} (1-u)} \textrm{Tr} ({\tilde F}_{ij})^2 \right] \ , \label{LIKKT_boson3}
\end{align}
where $\displaystyle {\tilde F}_{\mu \nu} = i \, [{\tilde A}_{\mu}, {\tilde A}_{\nu}]$.
Note that one obtains the Euclidean model at $u=1$.
Here we observe that the coefficients $e^{i \frac{\pi}{2}  (1-u)}$ and $e^{- i \frac{\pi}{2} (1-u)}$
have positive real parts for $0<u \leq 1$,
and hence the deformed model is well defined in that region.
Therefore we define the Lorentzian
model by taking the $u \rightarrow \!+0$ limit.

\begin{figure}
\centering
\includegraphics[scale=0.4]{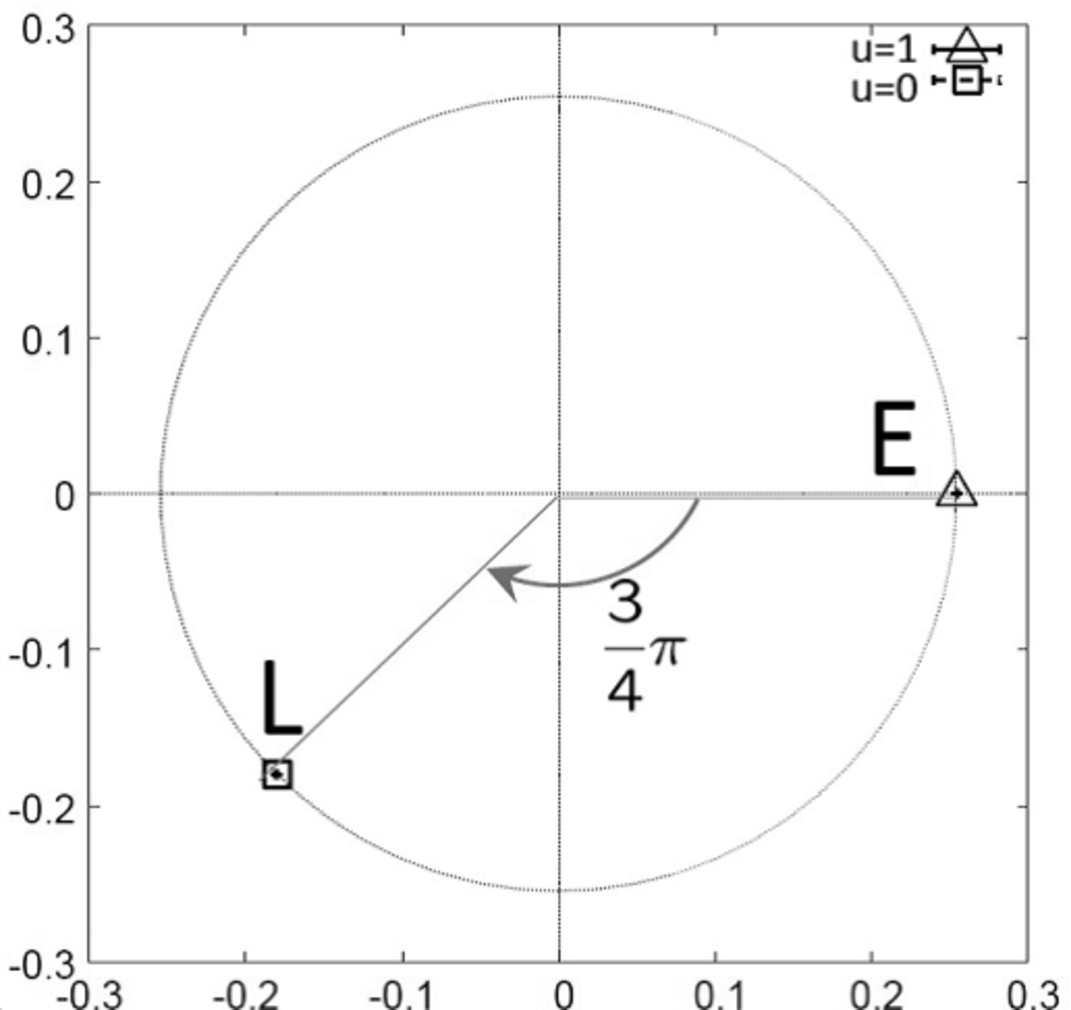} 
\includegraphics[scale=0.4]{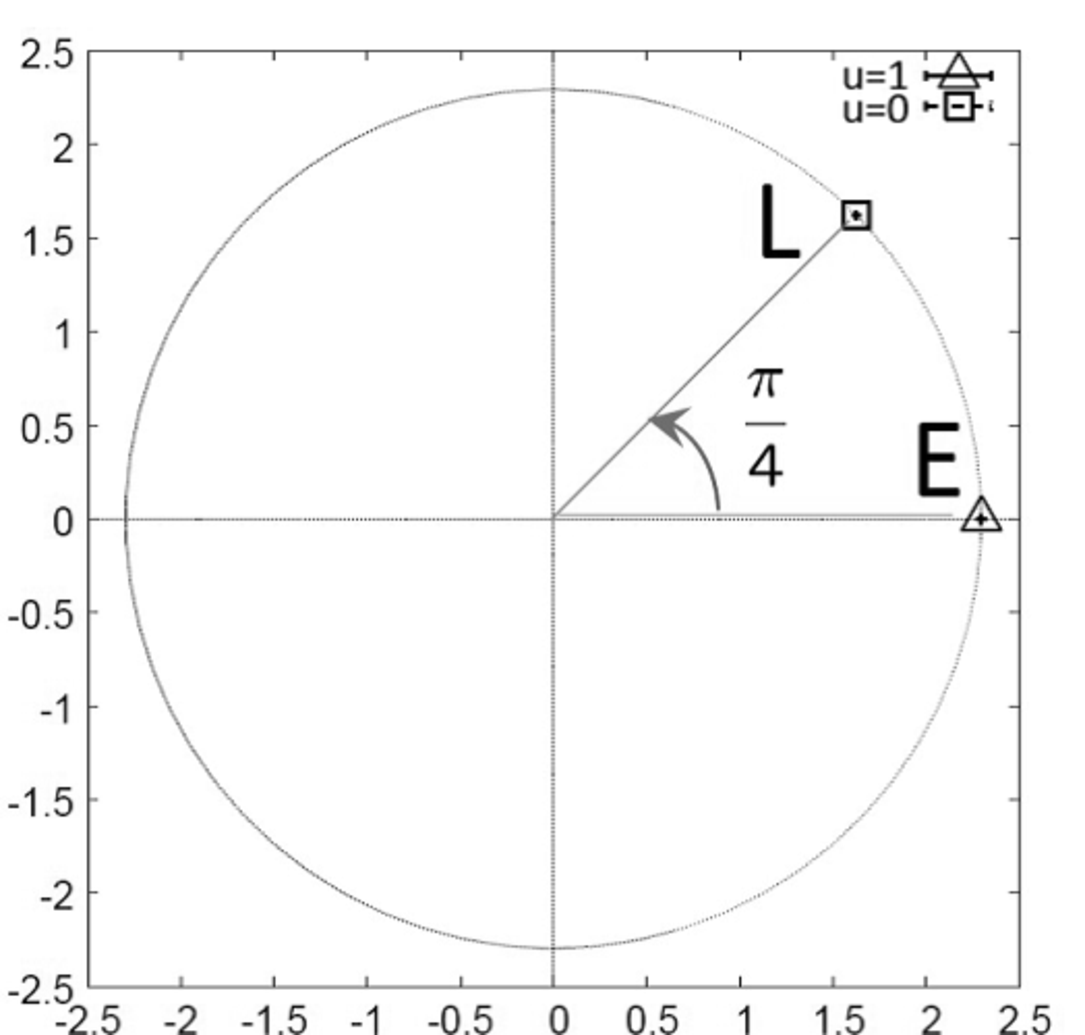} 
\caption{The expectation values of $\frac{1}{N}\Tr (A_0)^2$ (Left)
  and $\frac{1}{N}\Tr (A_i)^2$ (Right) are plotted in the complex plane
  for the Euclidean ({\sf E}) and Lorentzian ({\sf L}) models. 
  The relative complex phase for these quantities are
  $- \frac{3\pi}{4}$ and $\frac{\pi}{4}$, respectively,
  which agree with Eq.~\eqref{eq: TrAsq} \cite{Hatakeyama:2021ake,2201_13200}.
}
\label{fig: TrAisq}
\end{figure}

Let us note here that the observable
\begin{align}
  \langle {\cal O} (e^{- i \frac{3\pi}{8} u} {\tilde A}_0,
  \  e^{ i \frac{\pi}{8} u} {\tilde A}_i) \rangle_u
  \label{observable_u}
\end{align}
for the action \eqref{LIKKT_boson3} is independent of $u$
due to Cauchy's theorem.\footnote{We thank Yuhma Asano for pointing this out to us.}
The $u \rightarrow \! + 0$ limit and the $u=1$ case correspond to the Lorentzian and Euclidean
models, respectively.
In particular, we obtain the relationship
between the two models
with respect to
the expectation values of $\Tr (A_0)^2$ and $\Tr (A_i)^2$
as \cite{Hatakeyama:2021ake,2201_13200}
\begin{equation}
\label{eq: TrAsq}
\left\langle\frac{1}{N} \text{Tr} (A_0)^2\right\rangle_\mathrm{L}
=e^{-i\frac{3\pi}{4}}\left\langle\frac{1}{N} \text{Tr} (\tilde{A}_0)^2\right\rangle_\mathrm{E} \ ,
\quad
\left\langle\frac{1}{N} \text{Tr} (A_i)^2\right\rangle_\mathrm{L}
=e^{i\frac{\pi}{4}}\left\langle\frac{1}{N} \text{Tr} (\tilde{A}_i)^2\right\rangle_\mathrm{E} \ ,
\end{equation}
where $\expval{\ \cdot \ }_\mathrm{L}$ and $\expval{\ \cdot \ }_\mathrm{E}$
represent the expectation values in the Lorentzian and Euclidean models, respectively.
In Fig.~\ref{fig: TrAisq}, we plot the numerical results of
$\left\langle \frac{1}{N} \text{Tr} (A_0)^2 \right\rangle$
and $\left\langle \frac{1}{N} \text{Tr} (A_i)^2 \right\rangle$ obtained by applying the CLM,
described later in Section~\ref{Sec_CLM}, to the bosonic model \eqref{LIKKT_boson3}
without $\textrm{Pf} {\cal M}(A)$ for $N=128$.
We observe that
the relative complex phase for the two quantities
is $-\frac{3\pi}{4}$ and $\frac{\pi}{4}$, respectively,
which agree with Eq.~\eqref{eq: TrAsq}.



This implies, in particular, that
the emergent spacetime should be interpreted as Euclidean,
and moreover it is complex\footnote{This can happen even if $A_\mu$ are Hermitian
because of the complex weight $e^{iS_{\rm b}}$ in \eqref{eq: partition_func}.}
due to the phase factors in \eqref{eq: TrAsq}.
Note that
the equivalence holds also in the presence
of $\mathrm{Pf} \mathcal{M}(A)$,
which implies that all the results\footnote{For instance,
in the Euclidean version of the type IIB matrix model,
it has been found that the SO($10$)
rotational symmetry is spontaneously broken to SO($3$),
with three dimensions larger than the other
seven \cite{Nishimura:2011xy,Anagnostopoulos:2015gua,Anagnostopoulos:2020xai},
although the physical meaning of this is not clear.}
obtained for the Euclidean model \cite{Anagnostopoulos:2022dak}
can be translated trivially as the results for the Lorentzian model.

\subsection{Adding a Lorentz-invariant mass term}
In order to avoid the equivalence to the Euclidean model
and hence to define a genuinely Lorentzian model with a real emergent spacetime,
we add a Lorentz-invariant mass term to the action 
given by \cite{2201_13200,2205_04726,Anagnostopoulos:2022dak,2212_10127,2407_03491}
\begin{align}
  S_{\gamma} = - \ \frac{N}{2} \gamma \,\textrm{Tr} (A_{\mu} A^{\mu})
  = \frac{N}{2} \gamma\, \{ \textrm{Tr} (A_0)^{2} - \textrm{Tr} (A_{i})^2 \} \ .
  \label{Lorentzian_mass}
\end{align}
Setting ${\tilde S}_{\gamma} = -i S_{\gamma}$ so that it appears in the partition function
as $e^{-{\tilde S}_{\gamma}}$, we obtain
\begin{align}
 {\tilde S}_{\gamma} =
 \frac{N}{2} \gamma \, \{ e^{-\frac{i}{4} \pi(2+3u)} \textrm{Tr} ({\tilde A}_0)^{2}
 + e^{\frac{i}{4} \pi(2+u)} \textrm{Tr} ({\tilde A}_{i})^2 \} \ .
 \label{LIKKT_boson4} 
\end{align}
For $\gamma < 0$, the real parts of $\gamma \, e^{-\frac{i}{4}  \pi(2+3u)}$ and
$\gamma \, e^{\frac{i}{4} \pi(2+u)}$
are positive
for $0 < u \le 1$,
allowing the matrices to be rotated from the Lorentzian to the Euclidean model.
For $\gamma>0$, however, the real parts of these coefficients
are negative for $0 < u \le 1$, which implies that
the equivalence between the Lorentzian ($u \to \! +0$) and Euclidean  ($u=1$) models
does not hold any longer.
Thus we can avoid the equivalence to the Euclidean model for $\gamma > 0$.

The impact of the Lorentz-invariant mass term can be seen
already at the level of
classical solutions.
The classical equation of motion
of the Lorentzian type IIB matrix model with the mass term reads
\begin{equation}
[A^\nu, [A_\nu, A_\mu]]-\gamma A_\mu=0 \ .
\end{equation}
For $\gamma=0$, the solutions consist only of
simultaneously diagonalizable $A_\mu$
as proved in Appendix A of Ref.~\cite{Steinacker:2017vqw}.
In fact, one can generalize the proof\footnote{We thank Cheng-Tsung Wang for
pointing this out to us.} to show that the only solution 
for $\gamma<0$ is the trivial one $A_\mu = 0$.
For $\gamma > 0$, on the other hand,
it was found numerically \cite{Hatakeyama:2019jyw}
that nontrivial \emph{real} solutions do exist
and they typically represent an expanding spacetime with smooth space and time
although the dimension of space is left undetermined.
Thus it is conceivable that the Lorentz-invariant mass term \eqref{Lorentzian_mass}
with $\gamma>0$ 
changes the dynamics of the Lorentzian model drastically,
and that the large-$N$ limit followed by the $\gamma \rightarrow 0$ limit
provides its sensible definition.

Let us also note that $1/\gamma^2$ plays the role of $\hbar$
in quantum theory. This can be seen by rescaling the matrices
as $A_\mu \rightarrow \sqrt{|\gamma|} A_\mu$, which makes $\gamma^2$ an overall factor
of the whole bosonic action. Therefore, at sufficiently large $\gamma$, the dominant configurations
are expected to be close to a classical solution, which typically exhibits
expanding behavior \cite{Hatakeyama:2019jyw} for $\gamma>0$ as we reviewed above.
The purpose of this work is not only to confirm this expectation by explicit numerical simulations
but also to
see whether (3+1)-dimensional expanding spacetime can be realized
in the presence of quantum corrections including the effects of fermions.

\subsection{Extracting the time evolution}
\label{sec: time_evolution}

Here we discuss how we can
extract the time evolution from matrix configurations $A_\mu$
that contribute dominantly
to the partition function of the Lorentzian type IIB matrix model
with the mass term \eqref{Lorentzian_mass}.
For that, we first choose the SU($N$) basis in such a way that
$A_0$ is diagonal and its eigenvalues are arranged in the ascending order as \cite{Kim:2011cr}
\begin{align}
  A_0=\text{diag}(\alpha_1,\alpha_2,\dots,\alpha_N),\quad
    \alpha_1 <  \alpha_2 <  \dots < \alpha_N \ .
  \label{diagonal_gauge_A0}
\end{align}
In this basis, it turns out that
the spatial matrices $A_i$ typically exhibit a band-diagonal structure.\footnote{More precisely,
the band-diagonal structure appears when the emergent space exhibits an expanding behavior,
which can be understood from the bosonic part of the action \eqref{eq: action_b}.
When the eigenvalues of $A_i$ become large, the commutator squared terms
prefer simultaneously diagonalizable matrices.}
Namely, there exists an integer $n$ such that the modulus of the elements $(A_i)_{ab}$
with $|a-b|>n$ are significantly smaller than those
with $|a-b| \leq n$ \cite{Ito:2013qga,Ito:2013ywa,Ito:2015mxa,Ito:2015mem,Ito:2017rcr},
as we will see later by calculating
the quantity
\begin{align}
  {\cal A}_{ab} = \frac{1}{9} \sum_{i=1}^9
  (A_i)_{ab}  (A_i)_{ba}
  \qquad  (a,b=1,2,\dots,N) \ .
  \label{Apq_def}
\end{align}

This motivates the introduction of $n \times n$ matrices $\bar{A}_i(t)$
defined as \cite{Kim:2011cr} 
\begin{equation}
\label{eq: Abar}
\qty(\bar{A}_i)_{pq}(t_k) = \qty(A_i)_{k+p, k+q} \ ,
\end{equation}
where $p,q=1,2,\dots,n$ and $k=0,1,\dots,N-n$.
We define the time $t_{k}$ as
\begin{align}
  \label{eq: time}
  t_k &= \sum_{j=1}^{k-1} | \tilde{t}_{j+1} - \tilde{t}_j | \ , \quad \quad
  \tilde{t}_k = \frac{1}{n} \sum_{p=1}^{n} \langle \alpha_{k+p} \rangle \ .
\end{align}
The phase $\theta^{(\textrm{t})}_a$ of the time difference is defined by
\begin{align}
  \langle \Delta \alpha_a \rangle \propto e^{i \theta^{(\textrm{t})}_a}, \textrm{ where }
  \Delta \alpha_a  = \alpha_{a+1} - \alpha_a \label{time_diff}  \  .
\end{align}
The emergence of real time requires $\theta^{(\textrm{t})}_a = 0$, whereas
for $\gamma=0$, we obtain a Euclidean spacetime with
$\theta^{(\textrm{t})}_a = - \frac{3\pi}{8}$ due to \eqref{eq: TrAsq}.

Using ${\bar A}_i (t)$ defined in Eq.~\eqref{eq: Abar}, we introduce the following observables.
First the extent of space at time $t$ is defined as
\begin{align}
  R^2 (t) = \left\langle \frac{1}{n} \textrm{tr} \sum_{i=1}^{9} ({\bar A}_i(t))^2 \right\rangle
  = e^{2i \theta_{\rm s} (t)} |R^2 (t)| \  ,
  \label{R_sq_def}
\end{align}
where $\textrm{tr}$ denotes the trace over the $n \times n$ matrices.
The emergence of real space requires $\theta_\mathrm{s}(t) \sim 0$,
whereas for $\gamma=0$, we obtain a Euclidean spacetime
with $\theta_\mathrm{s}(t) = \frac{\pi}{8}$. (See \eqref{eq: TrAsq}.)

Next we define the ``moment of inertia tensor'' at time $t$ as 
\begin{align}
  T_{ij} (t) = \frac{1}{n} \textrm{tr} ({\bar A}_i (t) {\bar A}_j (t)) \ , \label{Tij_def}  
\end{align}
which is a $9 \times 9$ real symmetric matrix.
The eigenvalues $\lambda_{i}(t)$ of $T_{ij} (t)$ are arranged in the descending order as
$  \lambda_{1} (t) > \lambda_{2} (t) > \dots >  \lambda_{9} (t)$.
Note that $\langle \sum_{i=1}^9 \lambda_{i} (t) \rangle$ is equal
to $R^2(t)$ defined in Eq.~\eqref{R_sq_def}.
If $\textrm{Re} \langle \lambda_{1} (t) \rangle, \ \dots,$
$\textrm{Re} \langle \lambda_{d} (t) \rangle$ are larger than the rest,
it indicates spontaneous symmetry breaking from SO($9$) to SO($d$),
implying the dynamical generation of $d$-dimensional space.

When calculating $\langle \lambda_{i} (t) \rangle$
using the CLM, we encounter the issue that each eigenvalue $\lambda_{i} (t)$ breaks
holomorphicity as a function of $A_i$.
To circumvent this, we compute the coefficients $c_k$ from
$\det (z {\bf 1}_{9} - T (t))  = \sum_{k=0}^9 c_k  z^k$
for each configuration\footnote{The coefficients $c_k$ are calculated using Vieta's formula.
This method was suggested to us by L.L.~Salcedo.},
and calculate $\vev{c_k}$. Then, we solve the ninth-order equation
$\sum_{k=0}^9 \langle c_k \rangle z^k = 0$ for $z$ numerically,
whose solutions give a practical estimate of $\vev{\lambda_{i} (t)}$.
Error bars
can be computed using the standard jackknife method\footnote{Since the configurations
generated by the CLM are complexified,
$\lambda_{i} (t)$ calculated for each configuration become complex in general.
As an alternative method for evaluating $\vev{\lambda_{i} (t)}$, which cannot be fully justified,
we can order them by imposing
$ \textrm{Re} \lambda_{1} (t) > \textrm{Re} \lambda_{2} (t) > \dots
> \textrm{Re} \lambda_{9} (t)$
and calculate $\vev{\lambda_{i} (t)}$ naively.
The results obtained in this way are found to be the
same as the ones obtained by the method described above within statistical errors.}.

In order to investigate the smoothness of the emergent space,
we consider the eigenvalues of the $n\times n$ matrix \cite{Aoki:2019tby}
\begin{align}
 Q(t) = \sum_{i=1}^9 ({\bar A}_i (t))^2, \label{Qij}
\end{align}
whose eigenvalues provide the radial distribution
of the points in space.
The eigenvalues $q_p(t)$ ($p=1, 2, \dots , n$) of $Q(t)$
are arranged
in the descending order as
$ q_{1} (t) >  q_{2} (t) > \dots >  q_{n} (t)$.
The expectation values $\langle q_p(t) \rangle$
are calculated using the CLM in a way similar to the one used for $\langle \lambda_{i} (t)\rangle$.

\section{Application of the complex Langevin method (CLM)} \label{Sec_CLM}

Monte Carlo simulations of the Lorentzian type IIB matrix model are difficult
due to the sign problem.
Here we use
the CLM \cite{Parisi:1983mgm, Klauder:1983sp},
which has been successfully applied to various systems with the sign problem.
It involves formulating stochastic differential equations for complexified variables,
which can then be used to calculate expectation values under specific conditions.

\subsection{The basic formulation of the CLM}

Let us consider a model defined by the partition function $Z=\int dx\, w(x)$,
where $x \in \bbR^n$
and $w(x)$ is a complex-valued function.
In the CLM, we complexify the variables as $x \to z \in \bbC^n$,
and solve the complex Langevin equation
with the Langevin time $\sigma$ given by 
\begin{equation}
\label{Langevin_eq}
\frac{dz_k}{d\sigma}={\frac{1}{w(z)}\pdv{w(z)}{z_k}} +{\eta_k(\sigma)} \ .
\end{equation}
The first term on the right-hand side
is the ``drift term'',
while the second term represents the real Gaussian noise with a probability distribution
proportional to $\exp (- \frac{1}{4} \int d\sigma \sum_k \eta_k(\sigma)^2 )$.

The solution of the above equation corresponds to
the probability distribution $P(z, \bar{z} ;\sigma)$,
which satisfies the Fokker-Planck equation. Under certain conditions, this distribution
converges to a stationary distribution $P(z, \bar{z})$ as $\sigma\to\infty$,
and one can prove
$\int dz d\bar{z} \, {\cal O}(z) P(z, \bar{z}) = \int dx\, {\cal O}(x) w(x)$
for an observable ${\cal O}(x)$,
where the holomorphicity of $w(z)$ and ${\cal O}(z)$ plays a crucial role.
Then the expectation value of ${\cal O}(x)$ can be evaluated as
\begin{align}
  \vev{{\cal O}(x)}
  \simeq  \frac{1}{\sigma} \int_{\sigma_0}^{\sigma_0+\sigma} d\sigma' {\cal O}(z(\sigma')) \ ,
    \label{kna.001}
\end{align}
after some time $\sigma_0$ required for the evolution to reach equilibrium.
The above proof of justification may fail due to various reasons
such as large excursions in the imaginary direction, boundary terms, integration cycles,
and the influence of irrelevant saddle points in the complex plane.
Significant progress has been made in identifying and addressing these
pathologies \cite{Aarts:2009uq,Aarts:2009dg,Aarts:2011ax,Nishimura:2015pba,Nagata:2015uga,Nagata:2016vkn,Scherzer:2018hid,Scherzer:2019lrh,Seiler:2023kes,Hansen:2024kjm,Mandl:2025mav,Mandl:2026vdc},
leading to successful applications
of the CLM to numerous cases \cite{Sexty:2013ica,Fodor:2015doa,Sexty:2019vqx,Scherzer:2020kiu,Ito:2020mys,Tsutsui:2025jez,Asano:2025qfb}.
The CLM was also applied
to the Euclidean type IIB matrix model \cite{Anagnostopoulos:2017gos,Anagnostopoulos:2020xai,Anagnostopoulos:2022dak}, where results consistent with the
Gaussian expansion method \cite{Nishimura:2011xy,Aoyama:2010ry} were obtained.

When we apply the CLM to the Lorentzian type IIB matrix model,
we first choose the SU($N$) basis by imposing \eqref{diagonal_gauge_A0}
in order to make the extraction of time evolution possible in the subsequent analysis.
The crucial idea here is to perform a change of variables \cite{Nishimura:2019qal}
\begin{equation}
  \alpha_1=0, \quad  \alpha_a = \sum_{b=1}^{a-1} e^{\tau_b}
  \quad \text{for~} a = 2, \ldots , N \ ,
  \label{temporal_tau}
\end{equation}
and introduce new real variables $\tau_a$
so that the ordering \eqref{diagonal_gauge_A0} of $\alpha_a$ is enforced automatically.
Here we do not care about the tracelessness of $A_0$ since the trace part is decoupled
due to the shift symmetry $A_0 \to A_0 + \textrm{(const.)} {\bf 1}_N$.
Instead, we impose the tracelessness on $A_0$ after generating configurations
by applying the shift $A_0 \to A_0 - \left( \frac{1}{N} \textrm{Tr} A_0 \right) {\bf 1}_N$.
The gauge fixing \eqref{diagonal_gauge_A0} and
the change of variables \eqref{temporal_tau} introduce an extra term \cite{Nishimura:2019qal}
\begin{align}
  S_{\textrm{g.f.}} =
  - \log \prod_{1 \leq a <b \leq N} (\alpha_a - \alpha_b)^2 - \sum_{a=1}^{N-1} \tau_a
  \label{gf_term}
\end{align} 
in the action.

Thus the partition function of the model we investigate by the CLM is given by
\begin{align}
  Z_\mathrm{eff} = \int dA_i d \tau e^{-S_\mathrm{eff}}, \ \textrm{ where } S_\mathrm{eff}
  = -i (S_{\textrm{b}} + S_{\gamma}) + S_{\textrm{g.f.}} - \log \textrm{Pf} {\cal M} \ ,
  \label{total_Seff} 
\end{align}
where $S_{\textrm{b}}$, $S_{\gamma}$, $S_{\textrm{g.f.}}$ and $\textrm{Pf} {\cal M}$ are
defined by Eq.~\eqref{eq: action_b} with $g^2N=1$, Eq.~\eqref{Lorentzian_mass},
Eq.~\eqref{gf_term} and Eq.~\eqref{eq: partition_func}, respectively.
The dynamical variables are $\tau_a$, which are real, and $A_i$, which are
traceless Hermitian matrices.

In order to implement the CLM, we complexify $\tau_a$ and
allow $A_i$ to be general complex traceless matrices.
The complex Langevin equations are then given by
\begin{equation}
\frac{d\tau_a}{d\sigma}
=- \ {\pdv{S_\mathrm{eff}}{\tau_a}} +{\eta_a(\sigma)} \ ,
\quad
\frac{d(A_i)_{ab}}{d\sigma}
=- \ {\pdv{S_\mathrm{eff}}{(A_i)_{ba}}} +{(\eta_i)_{ab}(\sigma)} \ ,
\label{CL_eq_LIKKT}
\end{equation}
where $(\eta_i)_{ab} (\sigma)$ and $\eta_a (\sigma)$ are traceless Hermitian matrices
and real numbers, respectively, which are drawn from probability distributions
proportional to $\exp \left( - \frac{1}{4} \int d \sigma \,
\textrm{Tr} \, \eta_i (\sigma)^2 \right)$
and $\exp \left( - \frac{1}{4} \int d \sigma \, \eta_a (\sigma)^2 \right)$, respectively.
In practice, the equations (\ref{CL_eq_LIKKT}) are solved
by discretizing $\sigma$ using the second-order Runge-Kutta method as
\begin{align}
  (A_{i})_{ab} (\sigma +\Delta \sigma) &= (A_{i})_{ab} (\sigma)
  - \Delta \sigma \Biggl\{ \beta_1  \frac{\partial S_{\textrm{eff}}}{\partial (A_i)_{ba}}  (A(\sigma))  + \beta_2  \frac{\partial S_{\textrm{eff}}}{\partial (A_i)_{ba}}  (A'(\sigma))  \Biggr\}
  + \sqrt{\Delta \sigma} ({\tilde \eta}_{i})_{ab} (\sigma) \ ,
 \label{RK2_a1} \\
(A'_{i})_{ab} (\sigma) &= (A_{i})_{ab} (\sigma)
- \Delta \sigma  \frac{\partial S_{\textrm{eff}}}{\partial (A_i)_{ba}}  (A(\sigma)) 
 + \sqrt{\Delta \sigma} ({\tilde \eta}_{i})_{ab} (\sigma) \ ,
 \label{RK2_a2}
\end{align}
where $\Delta \sigma$ is the step size,
set to $\Delta \sigma=10^{-5}$.
The coefficients are chosen as
$ \beta_1 =\beta_2 =
\frac{1}{2} \left( 1 + \frac{N}{6} \Delta \sigma \right)$ \cite{Fukugita:1986tg}.
The factor $\sqrt{\Delta \sigma}$
arises from the normalization of the noise
terms ${\tilde \eta}_i (\sigma)$, which are traceless
Hermitian matrices following the probability
distribution proportional to $\exp \left( - \frac{1}{4} \sum_{\sigma}
\textrm{Tr} \, {\tilde \eta}_i (\sigma)^2 \right)$.
The same ${\tilde \eta}_i (\sigma)$ is used in Eqs.~(\ref{RK2_a1}) and (\ref{RK2_a2}).
The update of $\tau_a$ is performed similarly.

\subsection{Some techniques to make the CLM work}

To obtain reliable results equivalent to the path integral using the CLM,
we must address two potential issues.  The first is the
``excursion problem'' \cite{Aarts:2009uq,Aarts:2011ax},
which arises when $A_i$ deviate significantly from Hermitian matrices,
and $\tau_a$ become far from real numbers.
Gauge cooling \cite{Seiler:2012wz,Nagata:2015uga,1604_07717} is a technique commonly
employed to mitigate the excursion problem.
While there is no room for gauge cooling
since we have already fixed the gauge
as in \eqref{diagonal_gauge_A0}, 
we have not encountered this problem in the simulations reported in this work.

The second issue is the ``singular drift
problem'' \cite{Mollgaard:2013qra,Nishimura:2015pba,Aarts:2017vrv},
which arises when the drift terms in the complex Langevin equation \eqref{CL_eq_LIKKT}
become excessively large. 
It was found in Ref.~\cite{Nagata:2016vkn} that the CLM is justified when the probability
distributions of the drift norms
\begin{align}
  u_{\tau} = \sqrt{\frac{1}{N^3} \sum_{a=1}^{N-1}
    \left| {\pdv{S_\mathrm{eff}}{\tau_a}} \right|^2} \ , \
  u_{A} = \sqrt{\frac{1}{9N^3} \sum_{i=1}^9 \sum_{a,b=1}^N
    \left| {\pdv{S_\mathrm{eff}}{(A_i)_{ba}}} \right|^2}
  \label{drift_norms} 
\end{align}
exhibit exponential or faster decay.
All results presented in this paper satisfy this criterion.

\begin{figure}[t]
\centering
\includegraphics[width=0.49\textwidth]{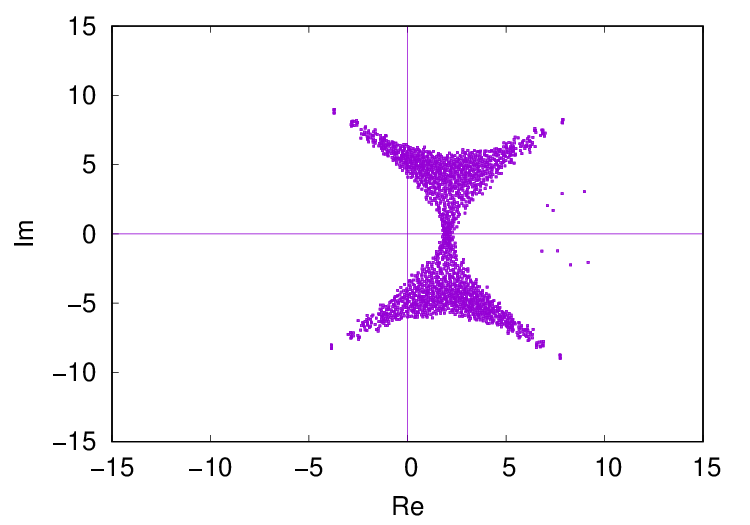}
\vspace*{1mm}
\caption{The eigenvalues of the matrix ${\cal M}$ obtained for $N=16$, $\gamma=6$,
  $\tilde{d}=6$, $\xi=18$ and $m_{\textrm{f}} = 2$,
  where the parameters $\tilde{d}$ and $\xi$ are introduced in Eq.\rf{Lorentzian_mass_xi}.}
\label{small_N_fig3}
\end{figure}

A naive application of the CLM to the type IIB matrix model does not work
due to the singular drift problem
caused by the near-zero eigenvalues of the matrix ${\cal M}(A)$
that appears in \eqref{eq: partition_func}.
Since the drift term includes the following contribution\footnote{This trace is calculated
for the $16 (N^2-1)\times 16(N^2-1)$ matrix
using the ``noisy estimator.'' See Appendix A of Ref.~\cite{Anagnostopoulos:2017gos} for details.
The noisy estimator involves solving the linear equation ${\cal M} \vec{x} = \vec{b}$,
where $\vec{x}$ and $\vec{b}$ are $16(N^2-1)$-dimensional vectors, using the conjugate
gradient (CG) method. Preconditioning with the input vector
$\vec{x}_0 = \frac{1}{m_{\textrm{f}}} \vec{b}$ reduces the required CG iterations by
approximately 10$\%$ compared to using the naive input vector $\vec{x}_0 =\vec{0}$.}
\begin{align}
  \frac{\partial}{\partial (A_i)_{ba}} \{ - \log \textrm{Pf} {\cal M} \} =
  - \ \frac{1}{2} \textrm{Tr} \left( \frac{\partial {\cal M}}
  {\partial (A_i)_{ba}} {\cal M}^{-1} \right) \ ,
  \label{singlar_drift_M}
\end{align}
near-zero eigenvalues of ${\cal M}$ can lead to the singular drift
problem.
Here we avoid this problem by adding a fermionic mass term \cite{Anagnostopoulos:2020xai}
\begin{align}
  S_{m_{\textrm{f}}} = - i \, m_{\textrm{f}}\, \frac{N}{2}\,   \textrm{Tr} {\bar \Psi}
  (\Gamma^7 (\Gamma^8)^{\dag} \Gamma^9) \Psi \ ,
  \label{fermionic_mass}
\end{align}
which modifies the matrix ${\cal M}(A)$ accordingly.
Note that the form of \eqref{fermionic_mass} is chosen
so that the Lorentz symmetry ${\rm SO}(9,1)$ is broken
minimally to ${\rm SO}(6,1)\times {\rm SO}(3)$.
One can also prove that the Pfaffian is real for $m_{\textrm{f}} \in \bbR$.
The effect of fermionic matrices diminishes with increasing $m_{\textrm{f}}$,
and in the $m_{\textrm{f}} \to \infty$ limit,
the fermionic matrices decouple completely,
resulting in the bosonic model.
For real $\mf$, this term shifts the eigenvalues of ${\cal M}(A)$ away from the origin
in the real direction as can be seen in Fig.~\ref{small_N_fig3}.

In fact, it turns out to be challenging to make $m_{\rm f}$ smaller than some value.
As we lower $\mf$, the computational cost increases due to slower convergence
of the conjugate gradient method used to evaluate Eq.~\eqref{singlar_drift_M}.
Furthermore, some of the eigenvalues of ${\cal M}(A)$ approach the origin,
causing the singular drift problem. The distributions of the drifts
in Eq.~\eqref{drift_norms} develop long tails that decay as a
power of $u_{\tau}$ or $u_{A}$,
failing to satisfy the criterion proposed in Ref.~\cite{Nagata:2016vkn},
and hence the CLM becomes unreliable.

Keeping $\mf$ sufficiently large to ensure reliable results suppresses the contributions from
fermion matrices.
In order to compensate for this suppression and to investigate the influence of SUSY
on the emergent spacetime, we enhance the fermionic contributions
by suppressing bosonic fluctuations.
Inspired by a BMN-type deformation \cite{Berenstein:2002jq} of the type IIB
matrix model\footnote{See Ref.~\cite{Kumar:2022giw,Kumar:2023nya,Hartnoll:2025ecj,Chou:2025rwy}
for Monte Carlo studies of the Euclidean model with this SUSY deformation.},
which preserves SUSY \cite{Bonelli:2002mb},
we modify the Lorentz-invariant mass term \eqref{Lorentzian_mass} as
\begin{equation}
    \label{Lorentzian_mass_xi}
    S_\gamma = \frac{1}{2}N\gamma
    \left( {\rm Tr}\left( A_0 \right)^2 -
    \sum_{i=1}^{\tilde{d}}{\rm Tr}\left( A_i \right)^2 -
    \xi \sum_{j=\tilde{d}+1}^9{\rm Tr}\left( A_j \right)^2  \right)\ ,
\end{equation}
where $\xi\ (\ge 1)$ is an additional parameter
introduced to suppress the fluctuations of $(9-\tilde{d})$ bosonic matrices.
This term breaks the Lorentz symmetry from SO($9,1$) to SO($\tilde{d},1$).
See Appendix \ref{sec:SUSY-deformation} for the relationship of
our deformation \eqref{fermionic_mass} and \eqref{Lorentzian_mass_xi}
to the SUSY deformation.

An additional technique to stabilize the simulation involves inserting
the procedure
\begin{align}
  A_i \mapsto \frac{A_i + \eta A_i^\dagger}{1+\eta}
  \quad \text{for~} i=1,\dots,9 \ ,
  \label{dynamical_stabilization}
\end{align}
after each Langevin step,
where the real positive parameter $\eta$ should be chosen as small as possible.
This technique is similar to the dynamical stabilization employed in lattice QCD
applications of the CLM although its justification is not rigorous \cite{Attanasio:2018rtq}.
Note that $\eta=1$ corresponds to enforcing Hermiticity on $A_i$.
In the following, we set $\eta=0.005$.

\section{Results of the complex Langevin simulations}
\label{results_sec}

In this section we present our numerical results obtained by the CLM.
First we show our results for the bosonic model, where the fermionic matrices are
omitted by hand. While we observe an expanding behavior, we find no SSB of the rotational
SO(9) symmetry.
Then we show our results for the SUSY model
with some deformation inspired by the SUSY deformation \cite{Bonelli:2002mb}.
We find a phase with (3+1)-dimensional expanding spacetime,
where only three out of nine directions
start to expand at some point in time. Unlike the previous
results \cite{Kim:2011cr,Ito:2013qga,Ito:2013ywa,Ito:2015mxa,Ito:2015mem,Ito:2017rcr}
with singular spatial structure \cite{Aoki:2019tby},
we find that the space is not only real but also smooth.


\subsection{Results for the bosonic model}
  \label{lorentz_boost_sec}

\begin{figure}
    \centering
    \includegraphics[height=12em]{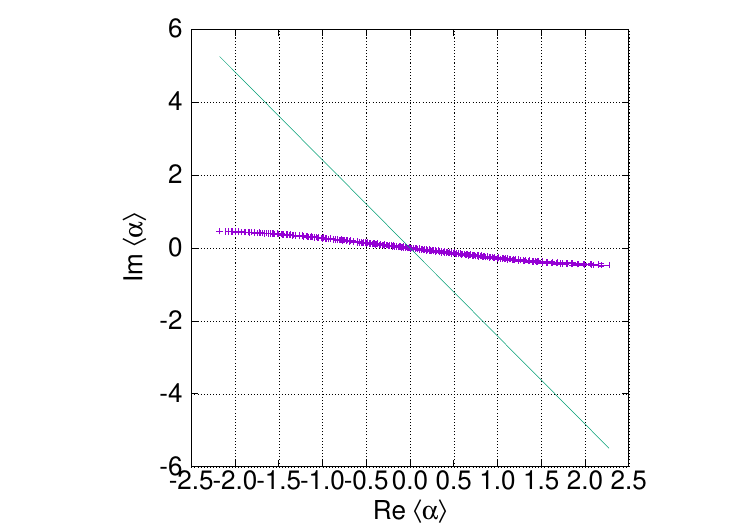}\hspace{2em}
    \includegraphics[height=12em]{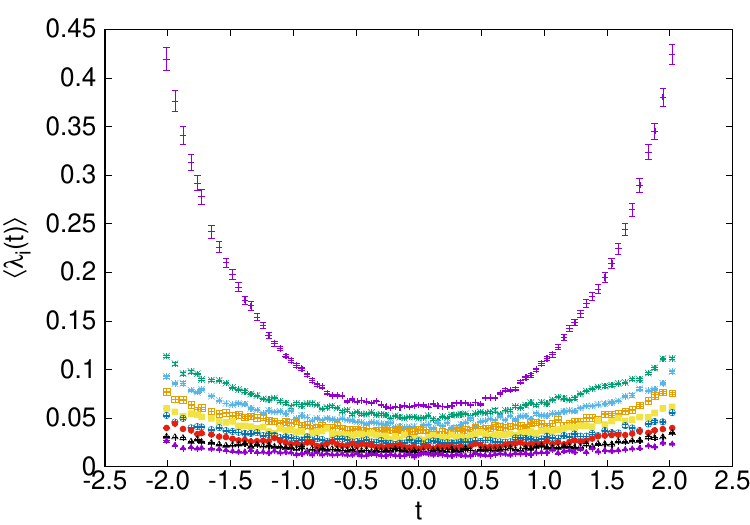}
    \caption{(Left) The expectation values $\vev{\alpha_a}$ of the eigenvalues
      of the temporal matrix $A_0$ are plotted in the complex plane for the bosonic model
      with $N=96$, $n=6$ and $\gamma=4$.
      The solid line represents the behavior \eqref{time_diff} expected for
      the original Lorentzian model ($\gamma=0$).
      (Right) The real parts of $\vev{\lambda_{i}(t)}$, the expectation values of
      the eigenvalues of $T_{ij}(t)$, are plotted against time for the same model.}
    \label{fig:alpha_Tij_before_LT}
\end{figure}

  Figure \ref{fig:alpha_Tij_before_LT} shows the expectation values of the eigenvalues $\alpha_a$
  of the temporal matrix $A_0$ and the eigenvalues $\lambda_{i}(t)$ of $T_{ij}(t)$
  for the bosonic model with the Lorentz-invariant mass term \eqref{Lorentzian_mass}
  for $N=96$, $n=6$ and $\gamma=4$, where $n$ is the block size used in \eqref{eq: Abar},
  and $\gamma$ is the coefficient of the mass term \eqref{Lorentzian_mass}.
  The initial configuration is chosen randomly by generating Gaussian variables
  for each element of the matrices.
  The near-horizontal distribution of $\vev{\alpha_a}$ at late times (large $|\alpha_a|$)
  indicates the emergence of real time.
We find that one of the nine eigenvalues of $T_{ij}(t)$ grows with $t$.
While this was misinterpreted as an evidence
for the emergence of an expanding (1+1)-dimensional spacetime in our earlier proceedings
articles \cite{2205_04726,2212_10127},
it was noticed in Ref.~\cite{2407_03491}
that it is actually an artifact of Lorentz boosts.
Since the model has the Lorentz symmetry, the configurations are Lorentz boosted
as the simulation proceeds, which implies that we have to be careful when
we interpret $A_0$ and $A_i$ as time and space, respectively, as we did in
Section \ref{sec: time_evolution}.

To assess the impact of the Lorentz boosts, we measure\footnote{Strictly speaking,
the quantity \eqref{trace_yi} and all the observables measured after the Lorentz transformation
discussed below are not holomorphic, which obscures their physical meaning in the CLM.
However, we consider that this is not a serious issue given that
the configurations $A_\mu$ obtained by complex Langevin simulations turn out
to be close to Hermitian. (See Figs.~\ref{fig:susy_alpha_eig_T} (Left) and
\ref{fig:susy_alpha_eig_Q}.)
Let us also note that, for $A_i$ in \eqref{trace_yi} and \eqref{tildeTij_def},
we actually use the Hermitian part $(A_i+A_i ^\dag)/2 $ of the configurations generated
by the CLM, which is justified only approximately.
Ideally, we should perform simulations
of the Lorentzian model after fixing the Lorentz symmetry \cite{Asano:2024def,2501_17798},
which is now on-going.}
\begin{align}
    X_i (t_a) = \frac{1}{n}\, s_i \,
   \textrm{tr}  {\bar A}'_{i} (t_a) 
  \, ,
  \qquad {\bar A}' _i (t) =  \sum_{j=1}^9 O_{ij}  {\bar A}_{j} (t) 
  \label{trace_yi} 
\end{align}
at each time  $t_a$ ($a=0,1,2,\ldots,N-n$),
where we do not sum over $i$, and $O$ is a $9 \times 9$ orthogonal matrix
defined by
\begin{align}
  \tilde{\Lambda} &= O \, \tilde{T} \, O^{\top} \ , \nn \\
  \tilde{T}_{ij}  &= \frac{1}{N} \textrm{Tr} (A_i A_j)  \ ,
  \label{tildeTij_def}  
\end{align}
where $\tilde{\Lambda}=
{\rm diag} ( \tilde{\lambda}_{1}  ,  \tilde{\lambda}_{2} ,  \dots ,  \tilde{\lambda}_{9} )$
is a diagonal matrix with 
$  \tilde{\lambda}_{1}  > \tilde{\lambda}_{2} > \dots >  \tilde{\lambda}_{9} $.
This quantity \eqref{trace_yi} 
represents the trace of each spatial
block matrix in the SO($9$) basis that diagonalizes
the moment of inertia tensor $\tilde{T}$ for the whole spacetime.
The sign $s_i$
is defined by $s_i = {\rm sgn}  \{ \textrm{tr} {\bar A}'_{i} (t_{N-n}) \} $,
associated with the latest time $t=t_{N-n}$.
This is needed
since otherwise the expectation value of \eqref{trace_yi} becomes trivially zero due to
the $\bbZ_2$ symmetry ($A_i \rightarrow -A_i$).

\begin{figure}
  \centering
  \includegraphics[height=12em]{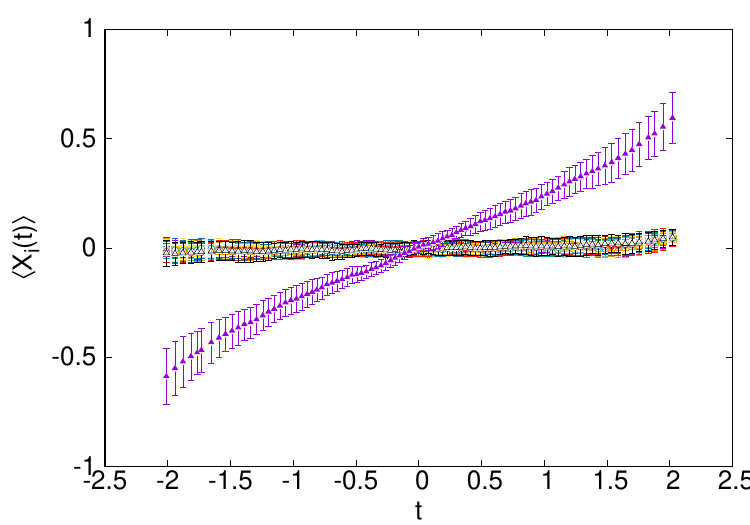}
    \hspace{2em}
    \includegraphics[height=12em]{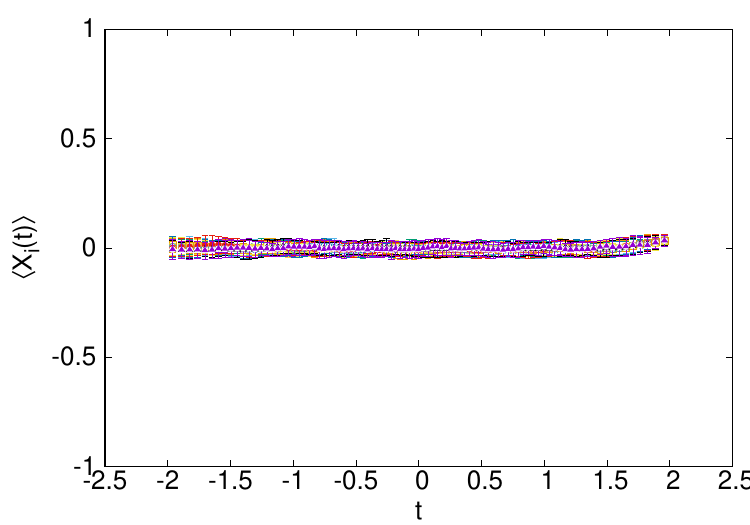}
    \caption{(Left) $ \langle X_i (t) \rangle$ defined by Eq.~\eqref{trace_yi}
      is plotted against time for the bosonic model
      with $N=96$, $n=6$ and $\gamma=4$ before the Lorentz transformations.
      Different colors correspond to different indices $i$.
      (Right) $ \langle X_i (t) \rangle$ is plotted against time for the same model
      after the Lorentz transformations.}
    \label{fig:trace_block_before_LT}
\end{figure}

In Fig.~\ref{fig:trace_block_before_LT} (Left), we plot
$ \langle X_i (t) \rangle$
for the configurations obtained in the simulation.
The result for the direction $i=1$ exhibits linear growth in time,
indicating that the obtained configurations
are Lorentz boosted.
Therefore, we must remove the artifact of Lorentz boosts to extract the
proper information about the emergent spacetime.

Figure \ref{fig:trace_block_before_LT} (Right) shows $\langle X_i (t) \rangle$
after the Lorentz transformations, which are determined in such a way that the
``center-of-mass motion'' is removed for each configuration.
(See Appendix \ref{sec:eliminating-effects-of-Lorentz=boosts} for the details of the procedure.)
Comparing this with Fig.~\ref{fig:trace_block_before_LT} (Left),
we find that the linear growth of the trace of the block matrices has been eliminated
by appropriate Lorentz transformations.


\begin{figure}[t]
    \centering
    \includegraphics[height=12em]{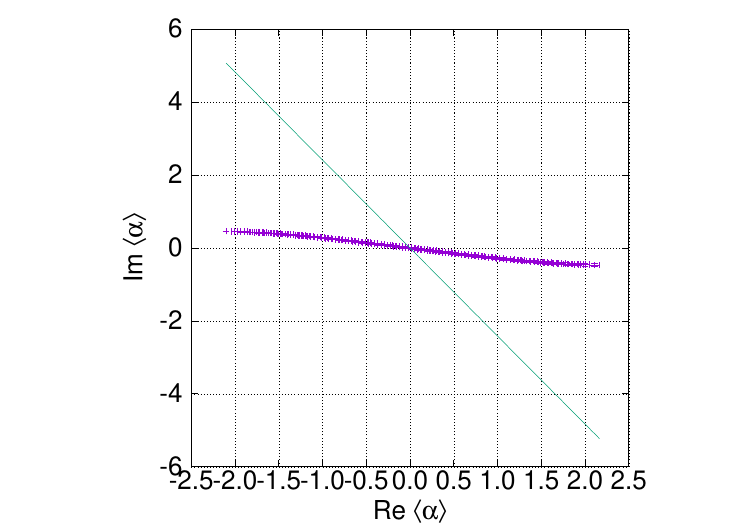}\hspace{2em}
    \includegraphics[height=12em]{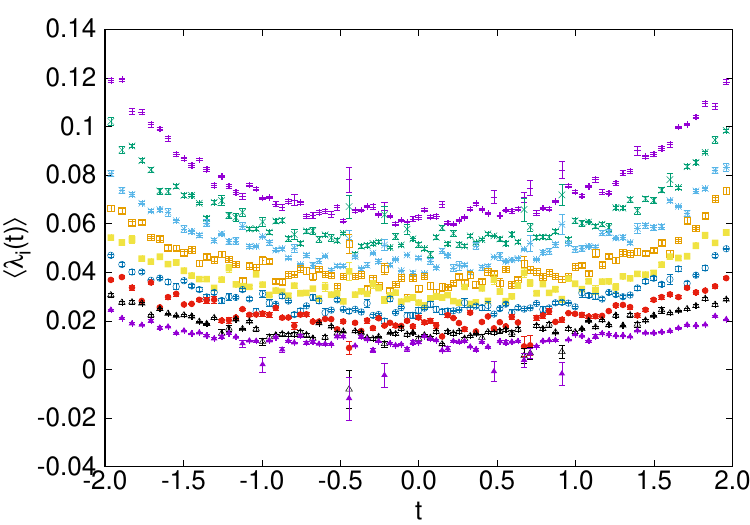}
    \caption{The same as Figure \ref{fig:alpha_Tij_before_LT} except that
      the results are obtained after the Lorentz transformations.}
    \label{fig:trace_block_Tij_after_LT}
\end{figure}

Figure \ref{fig:trace_block_Tij_after_LT} presents the expectation values of
the eigenvalues $\alpha_a$ of the temporal matrix $A_0$ (Left) and
the eigenvalues $\lambda_{i}(t)$ of $T_{ij}(t)$ (Right) after the Lorentz transformations.
This figure should be compared with Fig.~\ref{fig:alpha_Tij_before_LT},
which is obtained before the Lorentz transformations.
While the eigenvalue distribution of $\alpha_a$ remains largely unaffected by
the Lorentz transformations, the nine eigenvalues of $T_{ij}(t)$ converge significantly
after the transformations,
which indicates that the observed one-dimensional expansion is indeed
an artifact of Lorentz boosts.

\subsection{The effects of SUSY} \label{effect_SUSY}
In the previous section, we observed that the SO($9$) symmetry remains unbroken
in the bosonic model after removing the artifact of Lorentz boosts.
In this section, we perform similar analyses in the presence of the contributions from
fermionic matrices. In particular, we will see that
SUSY plays a crucial role in driving the expansion of three spatial directions.

\begin{figure}
    \centering
    \includegraphics[height=12em]{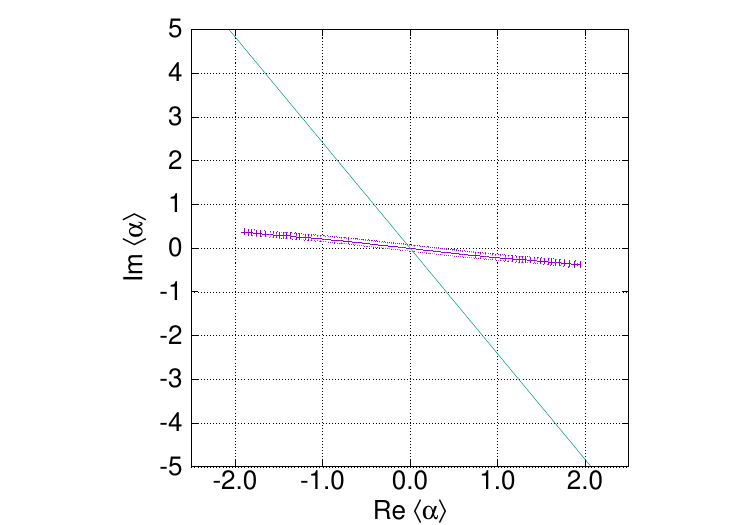}\hspace{2em}
    \includegraphics[height=12em]{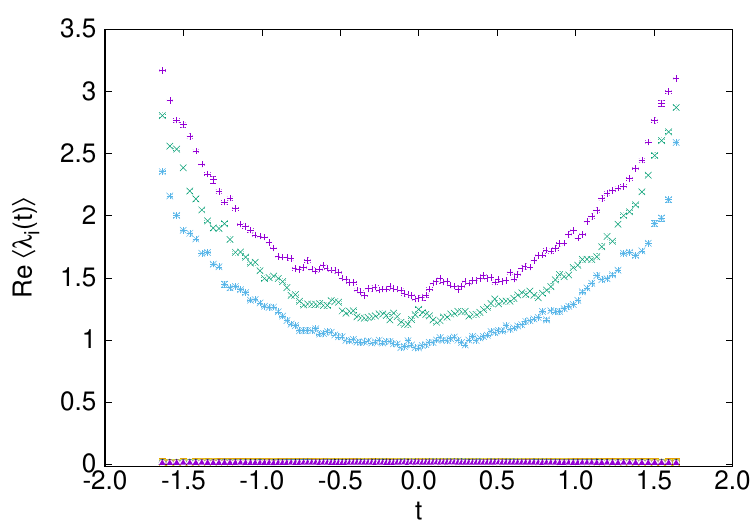}
    \caption{(Left) The expectation values $\vev{\alpha_a}$ of the eigenvalues
      of the temporal matrix $A_0$ are plotted in the complex plane for the bosonic model
      with $N=128$, $n=12$, $\gamma=4$, $\tilde d=3$ and $\xi=10$.
      The solid line represents the behavior \eqref{time_diff} expected for
      the original Lorentzian model ($\gamma=0$).
     (Right) The real parts of $\vev{\lambda_{i}(t)}$,
     the expectation values of the eigenvalues of  $T_{ij}(t)$, are plotted
     against time for the same model.
     The thermalized configuration is used as the initial configuration
     for the subsequent simulations.}
    \label{fig:init_alpha_eig_T}
\end{figure}

We begin with a simulation of the bosonic model using the modified 
mass term \eqref{Lorentzian_mass_xi}
with $\tilde{d}=3$ and $\xi=10$.
In Fig.~\ref{fig:init_alpha_eig_T}, we present the results of this simulation.
In the left panel, we show the eigenvalues of $A_0$.
In the right panel, we show
the expectation values $\vev{\lambda_{i}(t)}$
of the eigenvalues of $T_{ij}(t)$.
The thermalized configuration from this simulation is
then used as the initial configuration\footnote{\label{footnote:clm-weekpoint}Unfortunately, the CLM is not capable of
finding the most dominant saddle points. We have to prepare an initial configuration,
which is expected to be close to the most dominant saddle point.
In order to overcome this weakness of the method, we have to perform simulations
with the Lefschetz thimble method as has been done in Ref.~\cite{2501_17798} for $N=2$.
} for simulating the model with fermionic contributions
for $N=128$, $n=6$, $\gamma=4$,
where we use the SUSY-inspired deformation
given by the fermionic mass term \eqref{fermionic_mass}
with $m_{\textrm{f}} = \infty$\footnote{By $m_{\textrm{f}} = \infty$, we actually
mean the bosonic model, where $\textrm{Pf} {\cal M}$ is omitted by hand.}, $10 , 6$
and the modified bosonic mass term \eqref{Lorentzian_mass_xi}
with $\tilde{d}=5$ and $\xi=12$.
Below we present the results for this deformed model after applying
the Lorentz transformations to remove the artifact of Lorentz boosts.

\begin{figure}
    \centering
    \includegraphics[height=12em]{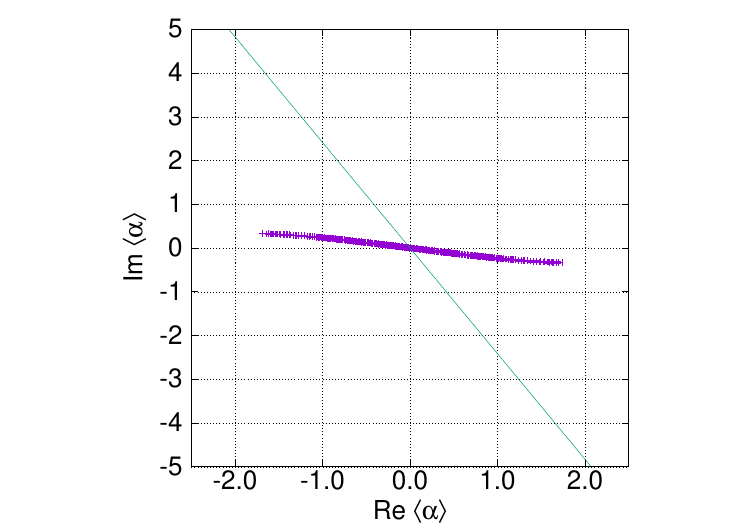}\hspace{2em}
    \includegraphics[height=12em]{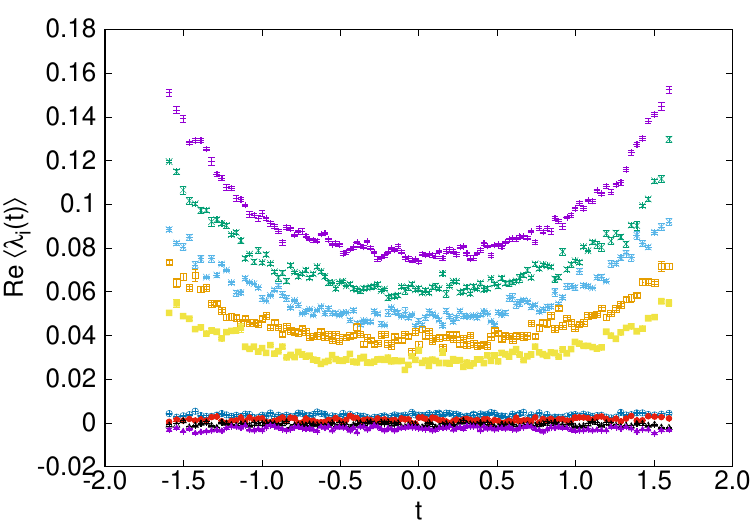}
    \includegraphics[height=12em]{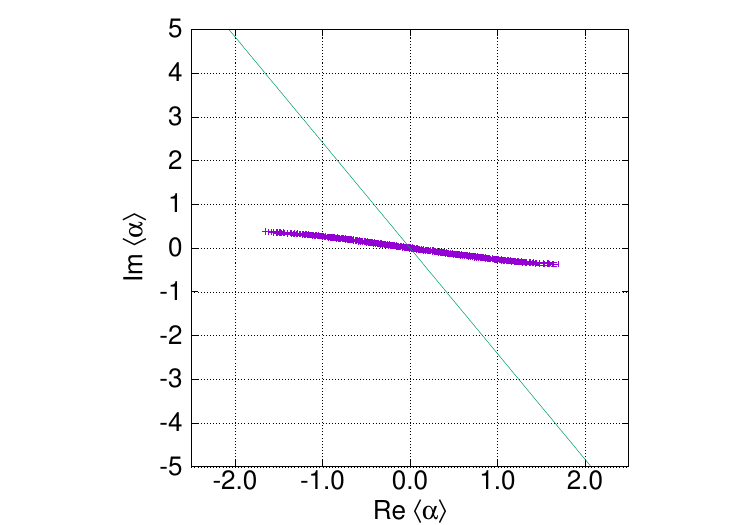}\hspace{2em}
    \includegraphics[height=12em]{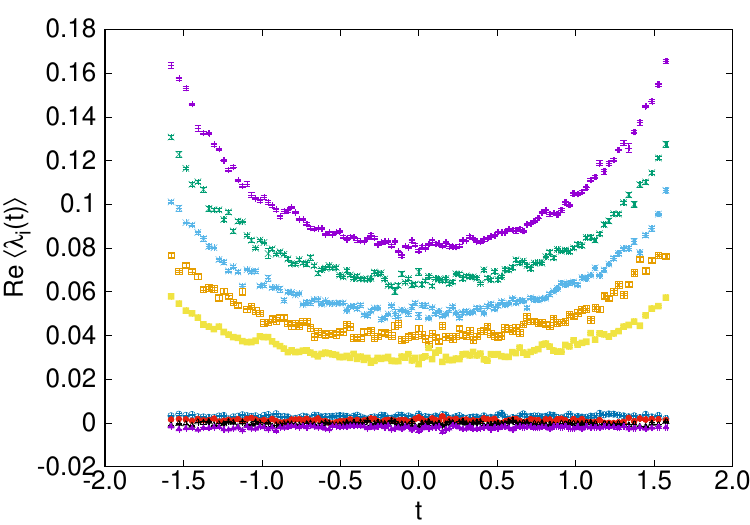}
    \includegraphics[height=12em]{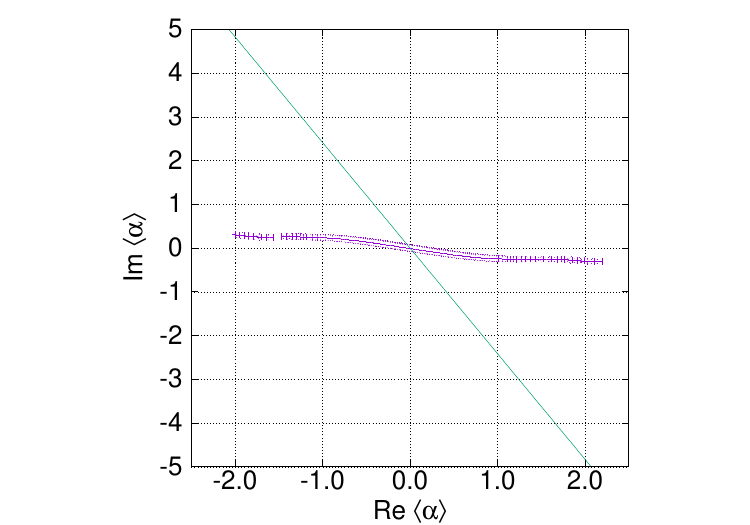}\hspace{2em}
    \includegraphics[height=12em]{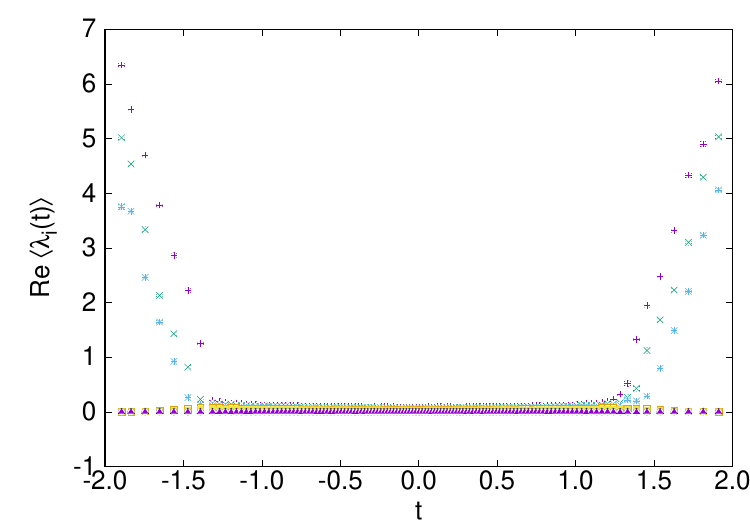}
    \caption{(Left) The expectation values $\vev{\alpha_a}$ of the eigenvalues
      of the temporal matrix $A_0$ are plotted in the complex plane for the bosonic model (Top),
      the model with fermionic contributions at $m_{\rm f}=10$ (Middle)
      and at $m_{\rm f}=6$ (Bottom), with $N=128$,  $n=6$, $\gamma=4$, $\tilde d=5$ and $\xi=12$.
      The solid line represents the behavior \eqref{time_diff} expected for
      the original Lorentzian model ($\gamma=0$).
      (Right) The real parts of $\vev{\lambda_{i}(t)}$, the expectation values of
      the eigenvalues of $T_{ij}(t)$, are plotted against time for the same models.}
    \label{fig:susy_alpha_eig_T}
\end{figure}

\begin{figure}
    \centering
    \includegraphics[height=12em]{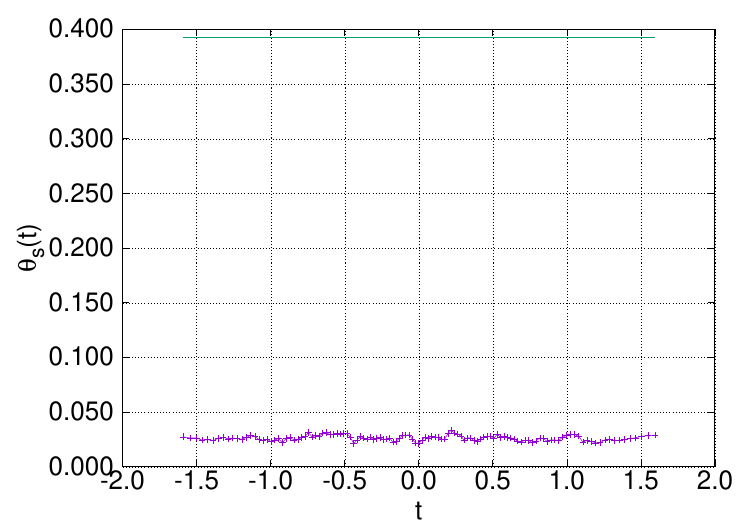}\hspace{2em}
    \includegraphics[height=12em]{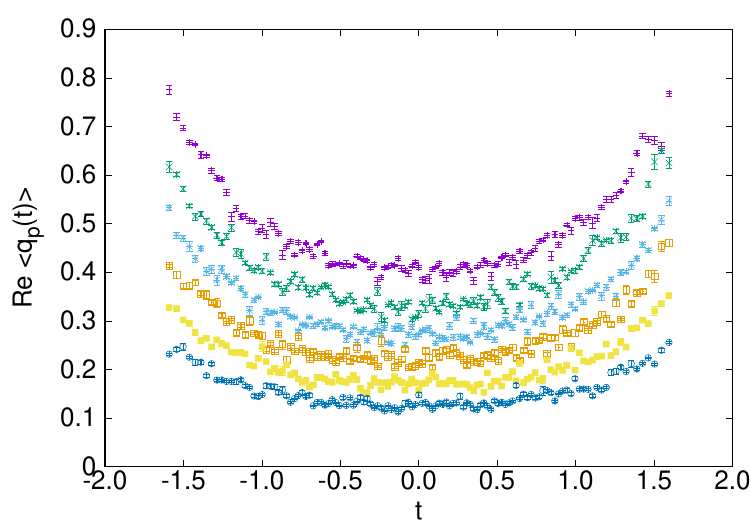}
    \includegraphics[height=12em]{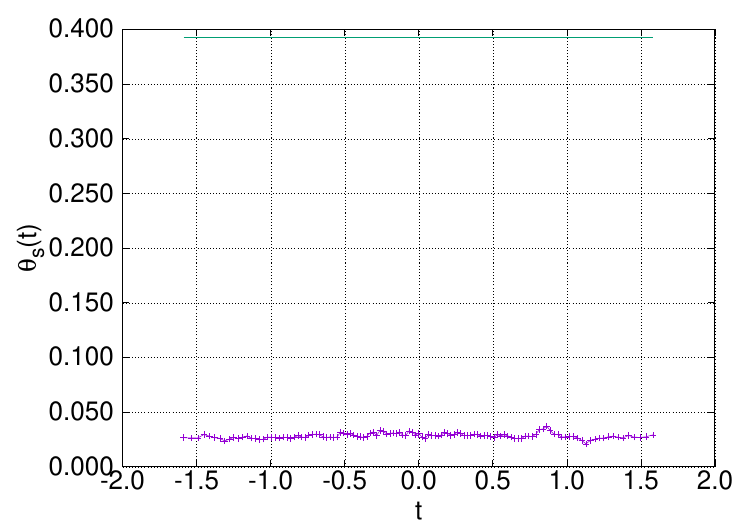}\hspace{2em}
    \includegraphics[height=12em]{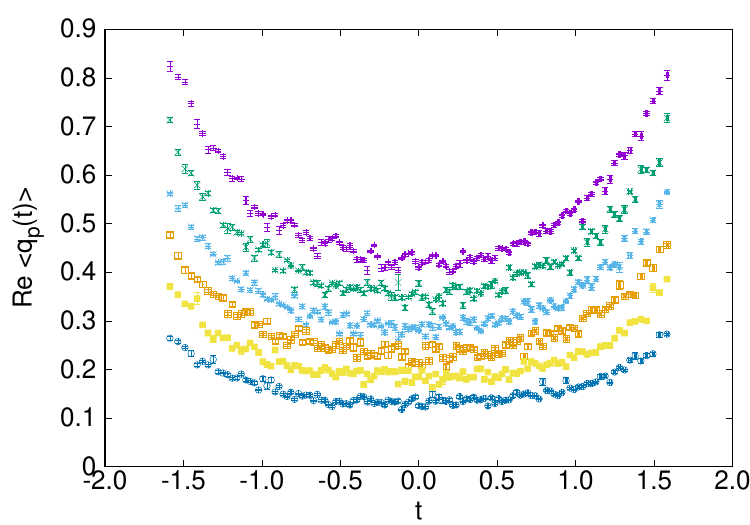}
    \includegraphics[height=12em]{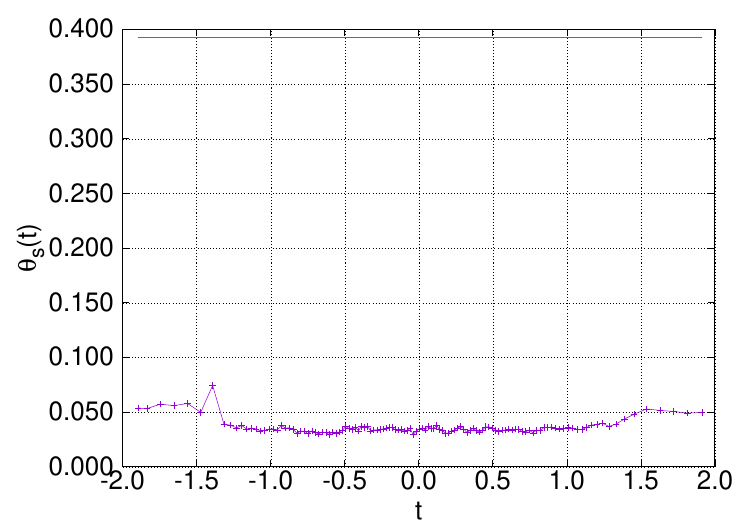}\hspace{2em}
    \includegraphics[height=12em]{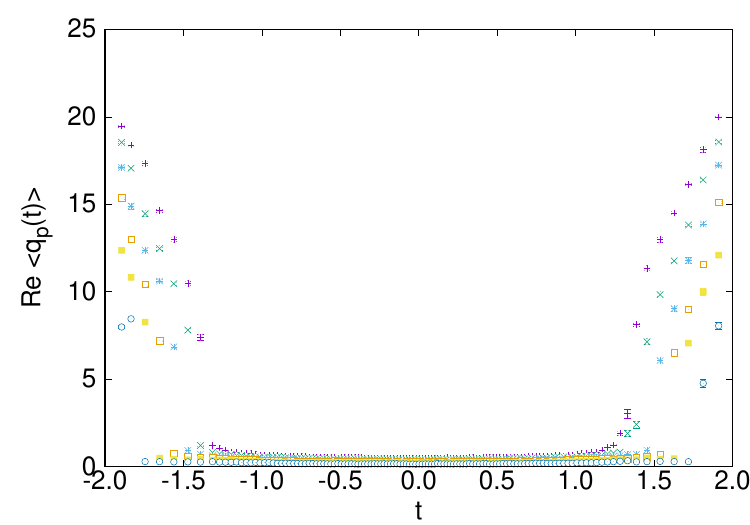}
    \caption{(Left) The phase of space $\theta_{\rm s}(t)$ is plotted for the bosonic model (Top)
      and for the model with fermionic contributions at $m_{\rm f}=10$ (Middle) and at $m_{\rm f}=6$ (Bottom), with $N=128$,  $n=6$, $\gamma=4$, $\tilde d=5$ and $\xi=12$.
      The horizontal line shows the value $\frac{\pi}{8}\sim 0.39$
      for the original Lorentzian model with $\gamma=0$,
      which is equivalent to the Euclidean model.
      (Right) The real parts of $\vev{q_p(t)}$,
      the expectation values of the eigenvalues of $Q(t)$, are plotted
      against time for the same models.}
    \label{fig:susy_alpha_eig_Q}
\end{figure}

In Fig.~\ref{fig:susy_alpha_eig_T}, we show the
results obtained for the $m_{\textrm{f}} = \infty , 10 , 6$ cases.
From the left panels, we find that
the eigenvalue distribution of $A_0$
becomes parallel to the real axis at late times (in particular, for the smallest $m_{\rm f}=6$),
indicating the emergence of real time in that regime.
From the right panels,
we find that
the SO($\tilde{d}$) rotational symmetry remains unbroken
for the $m_{\textrm{f}} = \infty$ and $m_{\textrm{f}} =10$ cases.
In contrast, in the $m_{\textrm{f}}=6$ case,
while the SO($\tilde{d}$) spatial rotational symmetry
appears to be preserved at early times close to the origin,
it is spontaneously broken to SO(3) at later times\footnote{If we start our simulation
from a random configuration as we did in the bosonic model, we obtain
results similar to those for $m_{\rm f} = 10$. See also footnote \ref{footnote:clm-weekpoint}.}.
Subsequently, only three eigenvalues exhibit rapid growth.
This result provides an evidence for a phase in which
an expanding (3+1)-dimensional spacetime emerges at late times.
As we increase $m_{\rm f}$ adiabatically from this configuration,
we find that simulations get unstable at $m_{\rm f} = 8$,
hinting at a phase boundary in that region.

In Fig.~\ref{fig:susy_alpha_eig_Q}, we show that the emergent space is
actually real and smooth for the three cases.
In the left panel, we present the results for
the phase $\theta_{\rm s}(t)$ of the space defined in \eqref{R_sq_def}.
The horizontal line shows the value $\frac{\pi}{8}\sim 0.39$
for the original Lorentzian model with $\gamma=0$,
which is equivalent to the Euclidean model.
The values obtained by simulations are an order of magnitude smaller
than this value.
In the right panel, we show the results for
the real parts of $\vev{q_p(t)}$, the expectation values of the eigenvalues of $Q(t)$.
We find that the eigenvalues are densely distributed unlike the case in which we find
singular structure described by Pauli matrices \cite{Aoki:2019tby}.

\begin{figure}
    \centering
    \includegraphics[width=0.49\hsize]{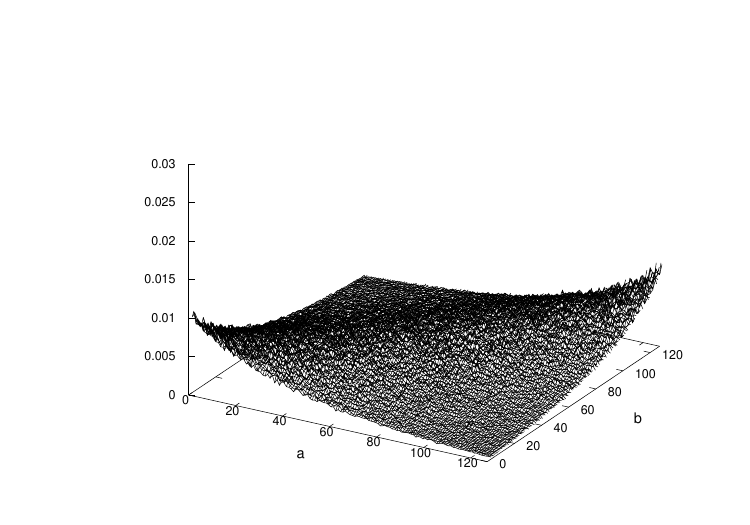}
    \includegraphics[width=0.49\hsize]{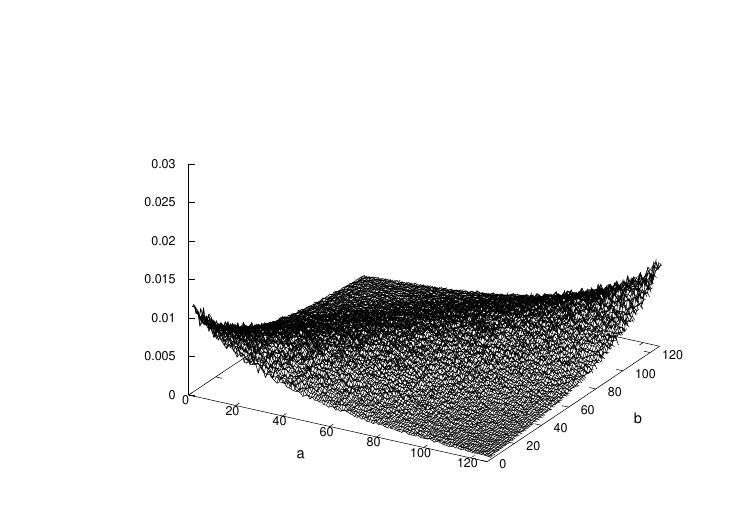}
    \includegraphics[width=0.49\hsize]{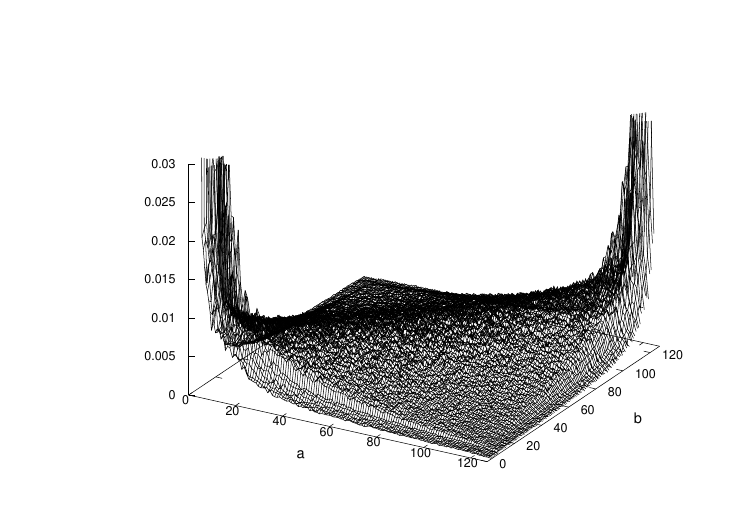}
    \caption{The quantity $|\mathcal{A}_{ab}|$, defined by Eq.~\eqref{Apq_def},
      is plotted for the bosonic model (Top-Left) and
      for the model with fermionic contributions at $m_{\rm f}=10$ (Top-Right)
      and at $m_{\rm f}=6$ (Bottom), with $N=128$, $n=6$, $\gamma=4$, $\tilde d=5$ and $\xi=12$.}
    \label{fig:band_diagonal_after_LT_SUSY}
\end{figure}

Figure \ref{fig:band_diagonal_after_LT_SUSY} shows the quantity $\mathcal{A}_{ab}$
defined by Eq.~\eqref{Apq_def} for the three cases,
illustrating the band-diagonal structure of the spatial matrices,
which plays a crucial role in extracting the time evolution as
we discussed in Section \ref{sec: time_evolution}.

\section{Summary and discussions}
\label{sec: conclusion}

In this paper, we investigated the Lorentzian version of the type IIB matrix model,
a proposed nonperturbative formulation of superstring theory,
employing the CLM to address the sign problem
with matrix size $N\leq 128$.
First we established the equivalence between the Lorentzian and Euclidean versions
of the type IIB matrix model under a contour deformation.
Consequently, one cannot obtain spacetime with Lorentzian signature from the original model,
which motivated us to introduce a Lorentz-invariant mass
term \eqref{Lorentzian_mass} in the action with $\gamma>0$.
Indeed, for sufficiently large $\gamma$,
the dominant configurations exhibit
the emergence of an expanding spacetime with the Lorentzian signature
as expected from the classical analysis \cite{Hatakeyama:2019jyw}.
However, we observed that the SO($9$) symmetry remains
unbroken in the bosonic model after removing the artifact of Lorentz boosts.

Adding the contributions from the fermionic matrices is not
straightforward in the CLM.
In order to avoid the singular drift problem of the CLM,
we added a fermionic mass term \eqref{fermionic_mass}
with the coefficient $m_{\textrm{f}}$ taken to be sufficiently large.
In the $m_{\textrm{f}} \to \infty$ limit, the fermionic matrices decouple and
one obtains the bosonic model.
Within the region of $m_{\textrm{f}}$ that allows us to avoid
the singular drift problem, we did not see any qualitative difference from
the bosonic model.
Inspired by the SUSY deformation \cite{Bonelli:2002mb} of the type IIB matrix model,
we therefore introduced some anisotropy in the bosonic mass term as
\eqref{Lorentzian_mass_xi}, where the parameter $\xi$ represents the anisotropy,
which enables us to control the quantum fluctuations of the bosonic matrices
preserving the ${\rm SO}(\tilde{d},1)$ Lorentz symmetry.
For $N=128$, $\gamma=4$, $\mf =6$, $\tilde d=5$ and $\xi=12$,
we found that the configurations obtained by the simulations
exhibit a prominent $(3+1)$-dimensional expanding behavior at late times,
which implies the SSB from SO($\tilde{d}$) to SO($3$).

Let us speculate on the mechanism for the emergence of an expanding $(3+1)$-dimensional spacetime.
As we saw in the studies of the Euclidean model \cite{Aoyama:2010ry,Nishimura:2011xy},
the volume of space tends to be constant when the SSB occurs in the type IIB matrix model.
Since the Pfaffian is a polynomial with respect to $A_\mu$, it is conceivable that
the modulus of the Pfaffian becomes larger for collapsed configurations due to the
increase of the linear extension of space caused by this constant-volume property.
On the other hand, when $\mf=0$, $\textrm{Pf} {\cal M}$ vanishes when $A_{\mu}$
is set to zero for all
but two indices \cite{hep-th/9803117,hep-th/0003223}.
This strongly disfavors
the emergence of spacetime with not greater than two spatial directions.
In this discussion, we assume that the temporal matrix $A_0$ can be ignored
since some of the spatial matrices become exponentially larger than the temporal matrix $A_0$
in the presence of the expanding behavior.

In our simulations, we left the Lorentz symmetry unfixed, which resulted in the
random Lorentz boosts during the simulation. While we were able to remove
the artifact of these Lorentz boosts by applying appropriate Lorentz transformations,
our analysis cannot be fully justified since it breaks holomorphicity required
for the correct application of the CLM.
We therefore think it important
to fix the Lorentz symmetry
nonperturbatively using the Faddeev-Popov procedure \cite{Asano:2024def,2501_17798}.
Simulation of this model by the CLM is on-going \cite{inprogress} at $N=128$,
and preliminary results suggest that our main conclusions remain the same.

In this work, we introduced the fermionic mass term to avoid the singular drift
problem in the CLM and added some anisotropy in the bosonic mass term to mimic the
SUSY deformation \cite{Bonelli:2002mb} of the type IIB matrix model.
However, there are some important differences from the SUSY deformed model
as we explain in Appendix \ref{sec:SUSY-deformation}.
Most importantly,
$\gamma>0$ requires pure imaginary $m_{\rm f}$ in the SUSY deformation,
whereas we have chosen $m_{\rm f}$ to be real, which is more effective in avoiding the
singular drift problem. (See Fig.~\ref{small_N_fig3}.)
Moreover, the cubic term in $A_\mu$ that has to be added
in the action for the SUSY deformation becomes purely imaginary for $\gamma>0$,
which may cause some instability due to unbounded action.
It is therefore quite nontrivial to simulate precisely the SUSY deformed model
and see whether the $(3+1)$-dimensional expanding behavior can still be obtained.
This is also under way \cite{inprogress}.

Last but not the least, the CLM is not powerful enough to sample all the relevant saddle
points and determine the most dominant one. For that, one needs to employ a yet another
promising approach to the sign problem, \emph{i.e.},
the Lefschetz thimble method \cite{2501_17798}.
While the computational cost is much larger than the CLM,
it is certainly worthwhile to develop the method further and clarify whether
the most dominant saddle point is given by the $(3+1)$-dimensional expanding spacetime.

\section*{Acknowledgments}
\vspace{-0.2cm}
We would like to thank Kohta Hatakeyama and Yuta Ito for their participation
at the earlier stage of this work. We are also grateful to Yuhma Asano and Naoyuki Yamamori
for valuable discussions and comments. T.~A. and A.~T. were supported in part by
Grant-in-Aid (Nos. 17K05425, and 18K03614, 21K03532, 25K07319, respectively)
from Japan Society for the Promotion of Science.
This research was supported by MEXT as ``Program for Promoting Researches
on the Supercomputer Fugaku'' (Simulation for basic science: approaching the new quantum era,
JPMXP1020230411) and JICFuS.  This work used computational resources of supercomputer
Fugaku provided by the RIKEN Center for Computational Science
(Project IDs: hp210165, hp220174, hp230207, hp240213, hp250224), and Oakbridge-CX provided
by the University of Tokyo (Project IDs: hp200106, hp200130, hp210094, hp220074, hp230149),
Grand Chariot provided by Hokkaido University (Project IDs: hp230149, hp240117),
and Flow provided by Nagoya University (Project ID: hp250140)
through the HPCI System Research Project.
Numerical computations were also carried out on Yukawa-21 at YITP in Kyoto University,
Flow in Nagoya University and PC clusters in KEK Computing Research Center.
This work was also supported by computational time granted by the Greek Research and
Technology Network (GRNET) in the National HPC facility ARIS,
under the project IDs LIIB2 and QCMM.

\appendix

\section{How to eliminate the artifact of Lorentz boosts}
\label{sec:eliminating-effects-of-Lorentz=boosts}

In this section, we explain how we determine
the Lorentz transformations that remove the artifact of Lorentz boosts.

The basic idea is to minimize the quantity
\begin{equation}
    \mathcal{T} = {\rm Tr}\left(A_0^\dagger A_0 \right)
\label{min_func}
\end{equation}
with respect to Lorentz transformations for each sampled configuration.
Practically, this can be achieved by iteratively
applying the (1+1)-dimensional Lorentz transformation
\begin{equation}
    \begin{pmatrix}
        A_0^\prime \\ A_i^\prime
    \end{pmatrix}
    =\begin{pmatrix}
        \cosh{\sigma} & \sinh{\sigma} \\ \sinh{\sigma} & \cosh{\sigma}
    \end{pmatrix}
    \begin{pmatrix}
        A_0 \\ A_i
    \end{pmatrix}\, ,
\end{equation}
where $i=1,2,...,\tilde{d}$ and $\sigma$ is a real parameter,
to minimize the quantity (\ref{min_func}) with respect to $\sigma$ at each step.\footnote{Note
that the Lorentz symmetry ${\rm SO}(9,1)$ is broken down to ${\rm SO}(\tilde{d},1)$
in the deformed model, where $\tilde{d}=5$ is chosen in our simulation.
Hence the Lorentz transformation is performed with $\tilde{d}=5$ in that case.}
To determine $\sigma$  at each step, we substitute  $A_0^\prime$ into Eq.~(\ref{min_func}),
obtaining
\begin{equation}
  \mathcal{T}^\prime = \cosh^2{\sigma}~ {\rm Tr}
  \left( A_0^\dagger A_0 \right) + \sinh^2{\sigma}~ {\rm Tr}
  \left( A_i^\dagger A_i \right) +
  2\cosh{\sigma}\sinh{\sigma}~ {\rm Re}~ {\rm Tr} \left( A_0^\dagger A_i \right) \  .
\end{equation}
Thus the problem reduces to minimizing
\begin{equation}
    f(x) = a\cosh{x} + b\sinh{x} \ ,
\end{equation}
where  $x=2\sigma$ and 
\begin{equation}
    a = {\rm Tr} \left( A_0^\dagger A_0 \right) + {\rm Tr} \left( A_i^\dagger A_i \right) \ ,\quad 
    b = 2~ {\rm Re}~ {\rm Tr} \left( A_0^\dagger A_i \right)\  .
\end{equation}
The minimum,  $\sqrt{a^2-b^2}$, is achieved at  $x = \tanh^{-1}{\left(-b/a\right)}$,
where $\left| b/a \right| < 1$,
which follows from the inequality
\begin{equation}
    {\rm Tr}\left(A_0 \pm A_i\right)^\dagger\left(A_0 \pm A_i\right) \ge 0\  .
\end{equation}

After the Lorentz transformation, the matrix $A_0$ is no longer diagonal.
Therefore, we redefine $\alpha_a$ by diagonalizing $A_0$ as
\begin{equation}
    A_0 \to U^{-1} A_0 U \equiv \tilde{A}_0\, ,
\end{equation}
where $U$ is a general complex matrix\footnote{Let us recall that the configurations $A_\mu$
generated by complex Langevin simulations are not Hermitian in general.}, and  $\tilde{A}_0 =
{\rm diag}(\alpha_1,\ \alpha_2,\ \dots,\ \alpha_N)$ is a complex diagonal matrix
with the ordering
\begin{equation}
    {\rm Re}~\alpha_1 < {\rm Re}~\alpha_2 < \ldots < {\rm Re}~\alpha_N\, .
\end{equation}
We also transform the spatial matrices  $A_i$ accordingly as
\begin{equation}
    A_i \to U^{-1} A_i U\, .
\end{equation}

\section{Relationship to the SUSY deformation}
\label{sec:SUSY-deformation}

In this section, we discuss a connection between the deformations
(\ref{fermionic_mass}) and (\ref{Lorentzian_mass_xi})
used in simulating the Lorentzian model in this paper
and the SUSY deformation of the type IIB matrix model \cite{Bonelli:2002mb}.
Let us consider a deformed model $S'$ of the type IIB matrix model with the action
\begin{align}
S' = S + S_{\gamma} + S_{\textrm{Myers}} + S_{m_{\textrm{f}}}  \ ,  
\end{align}
where $S$ is the action of the original model defined in (\ref{eq: action}) and
the terms $S_{\gamma}$ and $S_{m_{\textrm{f}}}$ are given by
(\ref{Lorentzian_mass_xi}) and (\ref{fermionic_mass}), respectively.
In addition to these terms, we also introduce the so-called Myers term
\begin{align}
S_{\textrm{Myers}} &= -iN\zeta\mbox{Tr} \left(A_7[A_8,A_9] \right)   \ .
\end{align}

By setting $\tilde{d}=6$, $\xi = 3$ and
\begin{align}
&\gamma = -\frac{\mu^2}{32} \, , \qquad
\zeta = \mu   \, ,  \qquad
m_{\textrm{f}}=\frac{\mu}{4} \, 
\nonumber
\end{align}
with $\mu$ being a real parameter,
one obtains the SUSY deformation of the type IIB matrix model \cite{Bonelli:2002mb}.
In that case, $S'$ is indeed invariant under a deformed SUSY transformation
\begin{align}
    \delta A_{\mu} &= i \bar{\epsilon}\Gamma_{\mu} \psi  \ , \nonumber\\
    \delta \psi &= \frac{i}{2}[A_{\mu},A_{\nu}] \Gamma^{\mu\nu}\epsilon  
    +\frac{\mu}{8} \{ (\Gamma^{\mu})^\dagger\Gamma^7(\Gamma^8)^{\dag}\Gamma^9
    +2(\Gamma^7)^\dagger\Gamma^8(\Gamma^9)^\dagger\Gamma^{\mu} \}A_{\mu}\epsilon\ ,
\end{align}
where $\epsilon$ is a constant Grassmann-odd parameter transforming as a Majorana-Weyl spinor
in (9+1) dimensions and
$\Gamma^{\mu\nu}=-\{(\Gamma^\mu)^\dagger\Gamma^\nu- (\Gamma^\nu)^\dagger\Gamma^\mu\}/2$.

On the other hand, the model that we simulate in section \ref{effect_SUSY}
corresponds to $S'$
for $\tilde{d}=5$, $\zeta=0$ with $\gamma$, $\xi$ and $m_{\textrm{f}}$
being real positive parameters.
Thus, the deformation used in this work
can be viewed as a generalization of the SUSY deformation of the type IIB matrix model.

The Euclidean version of the SUSY deformed model
has attracted a lot of attention
due to the discovery of the
holographic dual solutions in type IIB
supergravity \cite{Hartnoll:2024csr,Komatsu:2024bop,Komatsu:2024ydh}.
See also Refs.~\cite{Kumar:2022giw,Kumar:2023nya,Hartnoll:2025ecj,Chou:2025rwy}
for Monte Carlo studies of the Euclidean SUSY deformed model.

\bibliography{ref_CLMIKKT}

@article{Hansen:2024kjm,
    author = "Hansen, Michael W. and Mandl, Michael and Seiler, Erhard and Sexty, D\'enes",
    title = "{Role of integration cycles in complex Langevin simulations}",
    eprint = "2412.17137",
    archivePrefix = "arXiv",
    primaryClass = "hep-lat",
    doi = "10.1103/PhysRevD.111.074502",
    journal = "Phys. Rev. D",
    volume = "111",
    number = "7",
    pages = "074502",
    year = "2025"
}

@article{Seiler:2023kes,
    author = "Seiler, Erhard and Sexty, D\'enes and Stamatescu, Ion-Olimpiu",
    title = "{Complex Langevin: Correctness criteria, boundary terms, and spectrum}",
    eprint = "2304.00563",
    archivePrefix = "arXiv",
    primaryClass = "hep-lat",
    doi = "10.1103/PhysRevD.109.014509",
    journal = "Phys. Rev. D",
    volume = "109",
    number = "1",
    pages = "014509",
    year = "2024"
}

@article{Scherzer:2018hid,
    author = "Scherzer, Manuel and Seiler, Erhard and Sexty, D\'enes and Stamatescu, Ion-Olimpiu",
    title = "{Complex Langevin and boundary terms}",
    eprint = "1808.05187",
    archivePrefix = "arXiv",
    primaryClass = "hep-lat",
    doi = "10.1103/PhysRevD.99.014512",
    journal = "Phys. Rev. D",
    volume = "99",
    number = "1",
    pages = "014512",
    year = "2019"
}

@article{Seiler:2012wz,
    author = "Seiler, Erhard and Sexty, Denes and Stamatescu, Ion-Olimpiu",
    title = "{Gauge cooling in complex Langevin for QCD with heavy quarks}",
    eprint = "1211.3709",
    archivePrefix = "arXiv",
    primaryClass = "hep-lat",
    reportNumber = "MPP-2012-150",
    doi = "10.1016/j.physletb.2013.04.062",
    journal = "Phys. Lett. B",
    volume = "723",
    pages = "213--216",
    year = "2013"
}

@article{Aarts:2009dg,
    author = "Aarts, Gert and James, Frank A. and Seiler, Erhard and Stamatescu, Ion-Olimpiu",
    title = "{Adaptive stepsize and instabilities in complex Langevin dynamics}",
    eprint = "0912.0617",
    archivePrefix = "arXiv",
    primaryClass = "hep-lat",
    doi = "10.1016/j.physletb.2010.03.012",
    journal = "Phys. Lett. B",
    volume = "687",
    pages = "154--159",
    year = "2010"
}

@article{Komatsu:2024bop,
    author = "Komatsu, Shota and Martina, Adrien and Penedones, Jo{\~a}o and Vuignier, Antoine and Zhao, Xiang",
    title = "{Einstein gravity from a matrix integral. Part I}",
    eprint = "2410.18173",
    archivePrefix = "arXiv",
    primaryClass = "hep-th",
    doi = "10.1007/JHEP12(2025)029",
    journal = "JHEP",
    volume = "12",
    pages = "029",
    year = "2025"
}

@article{Komatsu:2024ydh,
    author = "Komatsu, Shota and Martina, Adrien and Penedones, Joao and Vuignier, Antoine and Zhao, Xiang",
    title = "{Einstein gravity from a matrix integral. Part II}",
    eprint = "2411.18678",
    archivePrefix = "arXiv",
    primaryClass = "hep-th",
    doi = "10.1007/JHEP12(2025)030",
    journal = "JHEP",
    volume = "12",
    pages = "030",
    year = "2025"
}

@article{Hartnoll:2024csr,
    author = "Hartnoll, Sean A. and Liu, Jun",
    title = "{The polarised IKKT matrix model}",
    eprint = "2409.18706",
    archivePrefix = "arXiv",
    primaryClass = "hep-th",
    doi = "10.1007/JHEP03(2025)060",
    journal = "JHEP",
    volume = "03",
    pages = "060",
    year = "2025"
}

@article{Steinacker:2021yxt,
    author = "Steinacker, Harold C.",
    title = "{Gravity as a quantum effect on quantum space-time}",
    eprint = "2110.03936",
    archivePrefix = "arXiv",
    primaryClass = "hep-th",
    reportNumber = "UWTHPh-2021-17",
    doi = "10.1016/j.physletb.2022.136946",
    journal = "Phys. Lett. B",
    volume = "827",
    pages = "136946",
    year = "2022"
}

@article{Battista:2023glw,
    author = "Battista, Emmanuele and Steinacker, Harold C.",
    title = "{One-loop effective action of the IKKT model for cosmological backgrounds}",
    eprint = "2310.11126",
    archivePrefix = "arXiv",
    primaryClass = "hep-th",
    doi = "10.1007/JHEP01(2024)125",
    journal = "JHEP",
    volume = "01",
    pages = "125",
    year = "2024"
}

@article{Steinacker:2023myp,
    author = "Steinacker, Harold C.",
    title = "{One-loop effective action and emergent gravity on quantum spaces in the IKKT matrix model}",
    eprint = "2303.08012",
    archivePrefix = "arXiv",
    primaryClass = "hep-th",
    reportNumber = "UWThPh-2023-9",
    doi = "10.1007/JHEP05(2023)129",
    journal = "JHEP",
    volume = "05",
    pages = "129",
    year = "2023"
}

@article{Kumar:2023bxg,
    author = "Kumar, Kaushlendra and Steinacker, Harold C.",
    title = "{Modified Einstein equations from the 1-loop effective action of the IKKT model}",
    eprint = "2312.01317",
    archivePrefix = "arXiv",
    primaryClass = "hep-th",
    doi = "10.1088/1361-6382/ad6e4b",
    journal = "Class. Quant. Grav.",
    volume = "41",
    number = "18",
    pages = "185007",
    year = "2024"
}

@article{Manta:2024vol,
    author = "Manta, Alessandro and Steinacker, Harold C. and Tran, Tung",
    title = "{$ \mathfrak{hs} $-extended gravity from the IKKT matrix model}",
    eprint = "2411.02598",
    archivePrefix = "arXiv",
    primaryClass = "hep-th",
    doi = "10.1007/JHEP02(2025)031",
    journal = "JHEP",
    volume = "02",
    pages = "031",
    year = "2025"
}

@article{Manta:2025inq,
    author = "Manta, Alessandro and Steinacker, Harold C.",
    title = "{Minimal covariant quantum space-time}",
    eprint = "2502.02498",
    archivePrefix = "arXiv",
    primaryClass = "hep-th",
    reportNumber = "UWThPh 2025-4",
    doi = "10.1088/1751-8121/adcc6e",
    journal = "J. Phys. A",
    volume = "58",
    number = "17",
    pages = "175204",
    year = "2025"
}

@article{Manta:2025tcl,
    author = "Manta, Alessandro and Steinacker, Harold C.",
    title = "{Dynamical covariant quantum spacetime with fuzzy extra dimensions in the IKKT model}",
    eprint = "2509.24753",
    archivePrefix = "arXiv",
    primaryClass = "hep-th",
    doi = "10.1007/JHEP02(2026)062",
    journal = "JHEP",
    volume = "02",
    pages = "062",
    year = "2026"
}

@book{Steinacker:2024unq,
    author = "Steinacker, Harold C.",
    title = "{Quantum geometry, matrix theory, and gravity}",
    doi = "10.1017/9781009440776",
    isbn = "978-1-009-44077-6, 978-1-009-44078-3",
    publisher = "Cambridge University Press",
    year = "2024"
}

@article{Aarts:2011ax,
    author = "Aarts, Gert and James, Frank A. and Seiler, Erhard and Stamatescu, Ion-Olimpiu",
    title = "{Complex Langevin: Etiology and diagnostics of its main problem}",
    eprint = "1101.3270",
    archivePrefix = "arXiv",
    primaryClass = "hep-lat",
    reportNumber = "MPP-2011-3",
    doi = "10.1140/epjc/s10052-011-1756-5",
    journal = "Eur. Phys. J. C",
    volume = "71",
    pages = "1756",
    year = "2011"
}

@article{Aarts:2009uq,
    author = "Aarts, Gert and Seiler, Erhard and Stamatescu, Ion-Olimpiu",
    title = "{The complex Langevin method: When can it be trusted?}",
    eprint = "0912.3360",
    archivePrefix = "arXiv",
    primaryClass = "hep-lat",
    doi = "10.1103/PhysRevD.81.054508",
    journal = "Phys. Rev. D",
    volume = "81",
    pages = "054508",
    year = "2010"
}

@article{Klinkhamer:2022frp,
    author = "Klinkhamer, F. R.",
    title = "{Emergent gravity from the IIB matrix model and cancellation of a cosmological constant}",
    eprint = "2212.00709",
    archivePrefix = "arXiv",
    primaryClass = "hep-th",
    reportNumber = "KA-TP-28-2022",
    doi = "10.1088/1361-6382/accef5",
    journal = "Class. Quant. Grav.",
    volume = "40",
    number = "12",
    pages = "124001",
    year = "2023"
}

@article{Brahma:2022ikl,
    author = "Brahma, Suddhasattwa and Brandenberger, Robert and Laliberte, Samuel",
    title = "{BFSS matrix model cosmology: Progress and challenges}",
    eprint = "2210.07288",
    archivePrefix = "arXiv",
    primaryClass = "hep-th",
    doi = "10.3390/physics5010001",
    journal = "MDPI Physics",
    volume = "5",
    number = "1",
    pages = "1--10",
    year = "2023"
}

@article{Asano:2024def,
    author = "Asano, Yuhma and Nishimura, Jun and Piensuk, Worapat and Yamamori, Naoyuki",
    title = "{Defining the type IIB matrix model without breaking Lorentz symmetry}",
    eprint = "2404.14045",
    archivePrefix = "arXiv",
    primaryClass = "hep-th",
    reportNumber = "UTHEP-787, KEK-TH-2617",
    doi = "10.1103/PhysRevLett.134.041603",
    journal = "Phys. Rev. Lett.",
    volume = "134",
    number = "4",
    pages = "041603",
    year = "2025"
}

@article{Klinkhamer:2020xoi,
    author = "Klinkhamer, F. R.",
    title = "{IIB matrix model and regularized big bang}",
    eprint = "2009.06525",
    archivePrefix = "arXiv",
    primaryClass = "hep-th",
    reportNumber = "KA-TP-14-2020",
    doi = "10.1093/ptep/ptab059",
    journal = "PTEP",
    volume = "2021",
    number = "6",
    pages = "063",
    year = "2021"
}

@article{Brahma:2022dsd,
    author = "Brahma, Suddhasattwa and Brandenberger, Robert and Laliberte, Samuel",
    title = "{Emergent metric space-time from matrix theory}",
    eprint = "2206.12468",
    archivePrefix = "arXiv",
    primaryClass = "hep-th",
    doi = "10.1007/JHEP09(2022)031",
    journal = "JHEP",
    volume = "09",
    pages = "031",
    year = "2022"
}

@article{Brandenberger:2024ddi,
    author = "Brandenberger, Robert and Pasiecznik, Julia",
    title = "{Origin of the SO(9){\textrightarrow}SO(3){\texttimes}SO(6) symmetry breaking in the type IIB matrix model}",
    eprint = "2409.00254",
    archivePrefix = "arXiv",
    primaryClass = "hep-th",
    doi = "10.1103/bqzp-n28g",
    journal = "Phys. Rev. D",
    volume = "112",
    number = "2",
    pages = "026006",
    year = "2025"
}

@article{Brahma:2021tkh,
    author = "Brahma, Suddhasattwa and Brandenberger, Robert and Laliberte, Samuel",
    title = "{Emergent cosmology from matrix theory}",
    eprint = "2107.11512",
    archivePrefix = "arXiv",
    primaryClass = "hep-th",
    doi = "10.1007/JHEP03(2022)067",
    journal = "JHEP",
    volume = "03",
    pages = "067",
    year = "2022"
}

@article{Anagnostopoulos:2022dak,
    author = "Anagnostopoulos, Konstantinos N. and Azuma, Takehiro and Hatakeyama, Kohta and Hirasawa, Mitsuaki and Ito, Yuta and Nishimura, Jun and Papadoudis, Stratos Kovalkov and Tsuchiya, Asato",
    title = "{Progress in the numerical studies of the type IIB matrix model}",
    eprint = "2210.17537",
    archivePrefix = "arXiv",
    primaryClass = "hep-th",
    reportNumber = "KEK-TH-2470",
    doi = "10.1140/epjs/s11734-023-00849-x",
    journal = "Eur. Phys. J. ST",
    volume = "232",
    number = "23-24",
    pages = "3681--3695",
    year = "2023"
}

@article{Anagnostopoulos:2010ux,
    author = "Anagnostopoulos, Konstantinos N. and Azuma, Takehiro and Nishimura, Jun",
    title = "{A general approach to the sign problem: The factorization method with multiple observables}",
    eprint = "1009.4504",
    archivePrefix = "arXiv",
    primaryClass = "cond-mat.stat-mech",
    reportNumber = "KEK-TH-1399",
    doi = "10.1103/PhysRevD.83.054504",
    journal = "Phys. Rev. D",
    volume = "83",
    pages = "054504",
    year = "2011"
}

@article{Anagnostopoulos:2013xga,
    author = "Anagnostopoulos, Konstantinos N. and Azuma, Takehiro and Nishimura, Jun",
    title = "{Monte Carlo studies of the spontaneous rotational symmetry breaking in dimensionally reduced super Yang-Mills models}",
    eprint = "1306.6135",
    archivePrefix = "arXiv",
    primaryClass = "hep-th",
    reportNumber = "KEK-TH-1634",
    doi = "10.1007/JHEP11(2013)009",
    journal = "JHEP",
    volume = "11",
    pages = "009",
    year = "2013"
}

@article{Anagnostopoulos:2011cn,
    author = "Anagnostopoulos, Konstantinos N. and Azuma, Takehiro and Nishimura, Jun",
    title = "{A practical solution to the sign problem in a matrix model for dynamical compactification}",
    eprint = "1108.1534",
    archivePrefix = "arXiv",
    primaryClass = "hep-lat",
    reportNumber = "KEK-TH-1482",
    doi = "10.1007/JHEP10(2011)126",
    journal = "JHEP",
    volume = "10",
    pages = "126",
    year = "2011"
}

@article{Anagnostopoulos:2001yb,
    author = "Anagnostopoulos, K. N. and Nishimura, J.",
    title = "{New approach to the complex action problem and its application to a nonperturbative study of superstring theory}",
    eprint = "hep-th/0108041",
    archivePrefix = "arXiv",
    reportNumber = "NBI-HE-01-08",
    doi = "10.1103/PhysRevD.66.106008",
    journal = "Phys. Rev. D",
    volume = "66",
    pages = "106008",
    year = "2002"
}

@article{Aoki:1998bq,
    author = "Aoki, Hajime and Iso, Satoshi and Kawai, Hikaru and Kitazawa, Yoshihisa and Tsuchiya, Asato and Tada, Tsukasa",
    editor = "Iso, S. and Kawai, H. and Natsuume, M.",
    title = "{IIB matrix model}",
    eprint = "hep-th/9908038",
    archivePrefix = "arXiv",
    reportNumber = "KEK-TH-635",
    doi = "10.1143/PTPS.134.47",
    journal = "Prog. Theor. Phys. Suppl.",
    volume = "134",
    pages = "47--83",
    year = "1999"
}

@article{Aoyama:2010ry,
    author = "Aoyama, Tatsumi and Nishimura, Jun and Okubo, Toshiyuki",
    title = "{Spontaneous breaking of the rotational symmetry in dimensionally reduced super Yang-Mills models}",
    eprint = "1007.0883",
    archivePrefix = "arXiv",
    primaryClass = "hep-th",
    doi = "10.1143/PTP.125.537",
    journal = "Prog. Theor. Phys.",
    volume = "125",
    pages = "537--563",
    year = "2011"
}

@article{Nishimura:2004ts,
    author = "Nishimura, Jun and Okubo, Toshiyuki and Sugino, Fumihiko",
    title = "{Gaussian expansion analysis of a matrix model with the spontaneous breakdown of rotational symmetry}",
    eprint = "hep-th/0412194",
    archivePrefix = "arXiv",
    reportNumber = "KEK-TH-1003, DPNU-04-23, OIQP-04-08",
    doi = "10.1143/PTP.114.487",
    journal = "Prog. Theor. Phys.",
    volume = "114",
    pages = "487--508",
    year = "2005"
}

@article{Nishimura:2000wf,
    author = "Nishimura, Jun and Vernizzi, Graziano",
    title = "{Brane world from IIB matrices}",
    eprint = "hep-th/0007022",
    archivePrefix = "arXiv",
    reportNumber = "NBI-HE-00-31",
    doi = "10.1103/PhysRevLett.85.4664",
    journal = "Phys. Rev. Lett.",
    volume = "85",
    pages = "4664--4667",
    year = "2000"
}

@article{Anagnostopoulos:2015gua,
    author = "Anagnostopoulos, Konstantinos N. and Azuma, Takehiro and Nishimura, Jun",
    title = "{Monte Carlo studies of dynamical compactification of extra dimensions in a model of nonperturbative string theory}",
    eprint = "1509.05079",
    archivePrefix = "arXiv",
    primaryClass = "hep-lat",
    reportNumber = "KEK-TH-1860",
    doi = "10.22323/1.251.0307",
    journal = "PoS",
    volume = "LATTICE2015",
    pages = "307",
    year = "2016"
}

@article{Ambjorn:2000dx,
    author = "Ambjorn, Jan and Anagnostopoulos, K. N. and Bietenholz, Wolfgang and Hotta, T. and Nishimura, J.",
    title = "{Monte Carlo studies of the IIB matrix model at large N}",
    eprint = "hep-th/0005147",
    archivePrefix = "arXiv",
    reportNumber = "NBI-HE-00-24, NORDITA-2000-50-HE",
    doi = "10.1088/1126-6708/2000/07/011",
    journal = "JHEP",
    volume = "07",
    pages = "011",
    year = "2000"
}

@article{Nishimura:2011xy,
    author = "Nishimura, Jun and Okubo, Toshiyuki and Sugino, Fumihiko",
    title = "{Systematic study of the SO(10) symmetry breaking vacua in the matrix model for type IIB superstrings}",
    eprint = "1108.1293",
    archivePrefix = "arXiv",
    primaryClass = "hep-th",
    reportNumber = "KEK-TH-1483, OIQP-11-06",
    doi = "10.1007/JHEP10(2011)135",
    journal = "JHEP",
    volume = "10",
    pages = "135",
    year = "2011"
}

@article{Aoki:1998vn,
    author = "Aoki, Hajime and Iso, Satoshi and Kawai, Hikaru and Kitazawa, Yoshihisa and Tada, Tsukasa",
    title = "{Space-time structures from IIB matrix model}",
    eprint = "hep-th/9802085",
    archivePrefix = "arXiv",
    reportNumber = "KEK-TH-559, TIT-HEP-382",
    doi = "10.1143/PTP.99.713",
    journal = "Prog. Theor. Phys.",
    volume = "99",
    pages = "713--746",
    year = "1998"
}

@article{Ishibashi:1996xs,
      author         = "Ishibashi, Nobuyuki and Kawai, Hikaru and Kitazawa, Yoshihisa and
                        Tsuchiya, Asato",
      title          = "{A large N reduced model as superstring}",
      journal        = "Nucl. Phys.",
      volume         = "B498",
      year           = "1997",
      pages          = "467-491",
      doi            = "10.1016/S0550-3213(97)00290-3",
      eprint         = "hep-th/9612115",
      archivePrefix  = "arXiv",
      primaryClass   = "hep-th",
      reportNumber   = "KEK-TH-503",
      SLACcitation   = "%%CITATION = HEP-TH/9612115;%%"
}

@article{Kim:2011cr,
      author         = "Kim, Sang-Woo and Nishimura, Jun and Tsuchiya, Asato",
      title          = "{Expanding (3+1)-dimensional universe from a Lorentzian
                        matrix model for superstring theory in (9+1)-dimensions}",
      journal        = "Phys. Rev. Lett.",
      volume         = "108",
      year           = "2012",
      pages          = "011601",
      doi            = "10.1103/PhysRevLett.108.011601",
      eprint         = "1108.1540",
      archivePrefix  = "arXiv",
      primaryClass   = "hep-th",
      reportNumber   = "KEK-TH-1484, OU-HET-720-2011",
      SLACcitation   = "%%CITATION = ARXIV:1108.1540;%%"
}

@article{Aoki:2019tby,
    author = "Aoki, Toshihiro and Hirasawa, Mitsuaki and Ito, Yuta and Nishimura, Jun and Tsuchiya, Asato",
    title = "{On the structure of the emergent 3d expanding space in the Lorentzian type IIB matrix model}",
    eprint = "1904.05914",
    archivePrefix = "arXiv",
    primaryClass = "hep-th",
    reportNumber = "KEK-TH-2110",
    doi = "10.1093/ptep/ptz092",
    journal = "PTEP",
    volume = "2019",
    number = "9",
    pages = "093B03",
    year = "2019"
}

@article{Nishimura:2019qal,
    author = "Nishimura, Jun and Tsuchiya, Asato",
    title = "{Complex Langevin analysis of the space-time structure in the Lorentzian type IIB matrix model}",
    eprint = "1904.05919",
    archivePrefix = "arXiv",
    primaryClass = "hep-th",
    reportNumber = "KEK-TH-2119",
    doi = "10.1007/JHEP06(2019)077",
    journal = "JHEP",
    volume = "06",
    pages = "077",
    year = "2019"
}

@article{Kim:2011ts,
      author         = "Kim, Sang-Woo and Nishimura, Jun and Tsuchiya, Asato",
      title          = "{Expanding universe as a classical solution in the
                        Lorentzian matrix model for nonperturbative superstring
                        theory}",
      journal        = "Phys. Rev.",
      volume         = "D86",
      year           = "2012",
      pages          = "027901",
      doi            = "10.1103/PhysRevD.86.027901",
      eprint         = "1110.4803",
      archivePrefix  = "arXiv",
      primaryClass   = "hep-th",
      reportNumber   = "KEK-TH-1503, OU-HET-729-2011",
      SLACcitation   = "%%CITATION = ARXIV:1110.4803;%%"
}

@article{Ito:2013ywa,
      author         = "Ito, Yuta and Kim, Sang-Woo and Koizuka, Yuki and
                        Nishimura, Jun and Tsuchiya, Asato",
      title          = "{A renormalization group method for studying the early
                        universe in the Lorentzian IIB matrix model}",
      journal        = "PTEP",
      volume         = "2014",
      year           = "2014",
      number         = "8",
      pages          = "083B01",
      doi            = "10.1093/ptep/ptu101",
      eprint         = "1312.5415",
      archivePrefix  = "arXiv",
      primaryClass   = "hep-th",
      reportNumber   = "KEK-TH-1696, KIAS-P13068",
      SLACcitation   = "%%CITATION = ARXIV:1312.5415;%%"
}

@article{Ito:2015mxa,
      author         = "Ito, Yuta and Nishimura, Jun and Tsuchiya, Asato",
      title          = "{Power-law expansion of the Universe from the bosonic
                        Lorentzian type IIB matrix model}",
      journal        = "JHEP",
      volume         = "11",
      year           = "2015",
      pages          = "070",
      doi            = "10.1007/JHEP11(2015)070",
      eprint         = "1506.04795",
      archivePrefix  = "arXiv",
      primaryClass   = "hep-th",
      reportNumber   = "KEK-TH-1836",
      SLACcitation   = "%%CITATION = ARXIV:1506.04795;%%"
}

@article{Ito:2017rcr,
      author         = "Ito, Yuta and Nishimura, Jun and Tsuchiya, Asato",
      title          = "{Universality and the dynamical space-time dimensionality
                        in the Lorentzian type IIB matrix model}",
      journal        = "JHEP",
      volume         = "03",
      year           = "2017",
      pages          = "143",
      doi            = "10.1007/JHEP03(2017)143",
      eprint         = "1701.07783",
      archivePrefix  = "arXiv",
      primaryClass   = "hep-th",
      reportNumber   = "KEK-TH-1952",
      SLACcitation   = "%%CITATION = ARXIV:1701.07783;%%"
}

@article{Steinacker:2017vqw,
      author         = "Steinacker, Harold C.",
      title          = "{Cosmological space-times with resolved Big Bang in
                        Yang-Mills matrix models}",
      journal        = "JHEP",
      volume         = "02",
      year           = "2018",
      pages          = "033",
      doi            = "10.1007/JHEP02(2018)033",
      eprint         = "1709.10480",
      archivePrefix  = "arXiv",
      primaryClass   = "hep-th",
      reportNumber   = "UWTHPH-2017-31",
      SLACcitation   = "%%CITATION = ARXIV:1709.10480;%%"
}

@article{Kim:2012mw,
      author         = "Kim, Sang-Woo and Nishimura, Jun and Tsuchiya, Asato",
      title          = "{Late time behaviors of the expanding universe in the IIB
                        matrix model}",
      journal        = "JHEP",
      volume         = "10",
      year           = "2012",
      pages          = "147",
      doi            = "10.1007/JHEP10(2012)147",
      eprint         = "1208.0711",
      archivePrefix  = "arXiv",
      primaryClass   = "hep-th",
      reportNumber   = "OU-HET-752-2012, KEK-TH-1563",
      SLACcitation   = "%%CITATION = ARXIV:1208.0711;%%"
}

@article{Aoki:2014cya,
      author         = "Aoki, Hajime and Nishimura, Jun and Tsuchiya, Asato",
      title          = "{Realizing three generations of the Standard Model
                        fermions in the type IIB matrix model}",
      journal        = "JHEP",
      volume         = "05",
      year           = "2014",
      pages          = "131",
      doi            = "10.1007/JHEP05(2014)131",
      eprint         = "1401.7848",
      archivePrefix  = "arXiv",
      primaryClass   = "hep-th",
      reportNumber   = "SAGA-HE-280, KEK-TH-1701",
      SLACcitation   = "%%CITATION = ARXIV:1401.7848;%%"
}

@article{Parisi:1983mgm,
    author = "Parisi, G.",
    title = "{On complex probabilities}",
    doi = "10.1016/0370-2693(83)90525-7",
    journal = "Phys. Lett. B",
    volume = "131",
    pages = "393--395",
    year = "1983"
}

@article{Klauder:1983sp,
    author = "Klauder, John R.",
    title = "{Coherent state Langevin equations for canonical quantum systems with applications to the quantized Hall effect}",
    reportNumber = "Print-83-0902 (BTL)",
    doi = "10.1103/PhysRevA.29.2036",
    journal = "Phys. Rev. A",
    volume = "29",
    pages = "2036--2047",
    year = "1984"
}

@article{Anagnostopoulos:2017gos,
    author = "Anagnostopoulos, Konstantinos N. and Azuma, Takehiro and Ito, Yuta and Nishimura, Jun and Papadoudis, Stratos Kovalkov",
    title = "{Complex Langevin analysis of the spontaneous symmetry breaking in dimensionally reduced super Yang-Mills models}",
    eprint = "1712.07562",
    archivePrefix = "arXiv",
    primaryClass = "hep-lat",
    reportNumber = "KEK-TH-2023",
    doi = "10.1007/JHEP02(2018)151",
    journal = "JHEP",
    volume = "02",
    pages = "151",
    year = "2018"
}

@article{Anagnostopoulos:2020xai,
    author = "Anagnostopoulos, Konstantinos N. and Azuma, Takehiro and Ito, Yuta and Nishimura, Jun and Okubo, Toshiyuki and Kovalkov Papadoudis, Stratos",
    title = "{Complex Langevin analysis of the spontaneous breaking of 10D rotational symmetry in the Euclidean IKKT matrix model}",
    eprint = "2002.07410",
    archivePrefix = "arXiv",
    primaryClass = "hep-th",
    reportNumber = "KEK-TH-2184",
    doi = "10.1007/JHEP06(2020)069",
    journal = "JHEP",
    volume = "06",
    pages = "069",
    year = "2020"
}

@article{1604_07717,
    author = "Nagata, Keitaro and Nishimura, Jun and Shimasaki, Shinji",
    title = "{Gauge cooling for the singular-drift problem in the complex Langevin method - a test in Random Matrix Theory for finite density QCD}",
    eprint = "1604.07717",
    archivePrefix = "arXiv",
    primaryClass = "hep-lat",
    reportNumber = "KEK-TH-1854, KEK-CP-322",
    doi = "10.1007/JHEP07(2016)073",
    journal = "JHEP",
    volume = "07",
    pages = "073",
    year = "2016"
}

@article{Nagata:2016vkn,
    author = "Nagata, Keitaro and Nishimura, Jun and Shimasaki, Shinji",
    title = "{Argument for justification of the complex Langevin method and the condition for correct convergence}",
    eprint = "1606.07627",
    archivePrefix = "arXiv",
    primaryClass = "hep-lat",
    reportNumber = "KEK-TH-1911",
    doi = "10.1103/PhysRevD.94.114515",
    journal = "Phys. Rev. D",
    volume = "94",
    number = "11",
    pages = "114515",
    year = "2016"
}

@article{Hatakeyama:2019jyw,
    author = "Hatakeyama, Kohta and Matsumoto, Akira and Nishimura, Jun and Tsuchiya, Asato and Yosprakob, Atis",
    title = "{The emergence of expanding space\textendash{}time and intersecting D-branes from classical solutions in the Lorentzian type IIB matrix model}",
    eprint = "1911.08132",
    archivePrefix = "arXiv",
    primaryClass = "hep-th",
    reportNumber = "KEK-TH-2169",
    doi = "10.1093/ptep/ptaa042",
    journal = "PTEP",
    volume = "2020",
    number = "4",
    pages = "043B10",
    year = "2020"
}

@article{Ito:2013qga,
    author = "Ito, Yuta and Kim, Sang-Woo and Nishimura, Jun and Tsuchiya, Asato",
    title = "{Monte Carlo studies on the expanding behavior of the early universe in the Lorentzian type IIB matrix model}",
    eprint = "1311.5579",
    archivePrefix = "arXiv",
    primaryClass = "hep-lat",
    reportNumber = "KEK-TH-1685",
    doi = "10.22323/1.187.0341",
    journal = "PoS",
    volume = "LATTICE2013",
    pages = "341",
    year = "2014"
}

@article{Ito:2015mem,
    author = "Ito, Yuta and Nishimura, Jun and Tsuchiya, Asato",
    title = "{Large-scale computation of the exponentially expanding universe in a simplified Lorentzian type IIB matrix model}",
    eprint = "1512.01923",
    archivePrefix = "arXiv",
    primaryClass = "hep-lat",
    reportNumber = "KEK-TH-1879",
    doi = "10.22323/1.251.0243",
    journal = "PoS",
    volume = "LATTICE2015",
    pages = "243",
    year = "2016"
}

@article{Nishimura:2015pba,
    author = "Nishimura, Jun and Shimasaki, Shinji",
    title = "{New insights into the problem with a singular drift term
    in the complex Langevin method}",
    eprint = "1504.08359",
    archivePrefix = "arXiv",
    primaryClass = "hep-lat",
    reportNumber = "KEK-TH-1816",
    doi = "10.1103/PhysRevD.92.011501",
    journal = "Phys. Rev. D",
    volume = "92",
    number = "1",
    pages = "011501",
    year = "2015"
}

@article{Nagata:2015uga,
    author = "Nagata, Keitaro and Nishimura, Jun and Shimasaki, Shinji",
    title = "{Justification of the complex Langevin method with the gauge cooling procedure}",
    eprint = "1508.02377",
    archivePrefix = "arXiv",
    primaryClass = "hep-lat",
    reportNumber = "KEK-TH-1855",
    doi = "10.1093/ptep/ptv173",
    journal = "PTEP",
    volume = "2016",
    number = "1",
    pages = "013B01",
    year = "2016"
}

@article{Hatakeyama:2021ake,
    author = "Hatakeyama, Kohta and Anagnostopoulos, Konstantinos and Azuma, Takehiro and Hirasawa, Mitsuaki and Ito, Yuta and Nishimura, Jun and Papadoudis, Stratos and Tsuchiya, Asato",
    title = "{Relationship between the Euclidean and Lorentzian versions of the type IIB matrix model}",
    eprint = "2112.15368",
    archivePrefix = "arXiv",
    primaryClass = "hep-lat",
    reportNumber = "KEK-TH-2373",
    doi = "10.22323/1.396.0341",
    journal = "PoS",
    volume = "LATTICE2021",
    pages = "341",
    year = "2022"
}

@article{Fukugita:1986tg,
    author = "Fukugita, M. and Oyanagi, Y. and Ukawa, A.",
    title = "{Langevin simulation of the full QCD hadron mass spectrum on a lattice}",
    reportNumber = "RIFP-686",
    doi = "10.1103/PhysRevD.36.824",
    journal = "Phys. Rev. D",
    volume = "36",
    pages = "824",
    year = "1987"
}

@article{Attanasio:2018rtq,
    author = {Attanasio, Felipe and J\"ager, Benjamin},
    title = "{Dynamical stabilisation of complex Langevin simulations of QCD}",
    eprint = "1808.04400",
    archivePrefix = "arXiv",
    primaryClass = "hep-lat",
    reportNumber = "CP3-Origins-2018-30 DNRF90, NT@UW-18-05",
    doi = "10.1140/epjc/s10052-018-6512-7",
    journal = "Eur. Phys. J. C",
    volume = "79",
    number = "1",
    pages = "16",
    year = "2019"
}

@article{2201_13200,
    author = "Hatakeyama, Kohta and Anagnostopoulos, Konstantinos and Azuma, Takehiro and Hirasawa, Mitsuaki and Ito, Yuta and Nishimura, Jun and Papadoudis, Stratos and Tsuchiya, Asato",
    title = "{Complex Langevin studies of the emergent space-time in the type IIB matrix model}",
    eprint = "2201.13200",
    archivePrefix = "arXiv",
    primaryClass = "hep-th",
    reportNumber = "KEK-TH-2393",
    doi = "10.1142/9789811261633_0002",
    month = "1",
    year = "2022"
}

@article{2205_04726,
    author = "Nishimura, Jun",
    title = "{Signature change of the emergent space-time in the IKKT matrix model}",
    eprint = "2205.04726",
    archivePrefix = "arXiv",
    primaryClass = "hep-th",
    reportNumber = "KEK-TH-2425",
    doi = "10.22323/1.406.0255",
    journal = "PoS",
    volume = "CORFU2021",
    pages = "255",
    year = "2022"
}

@article{2212_10127,
    author = "Hirasawa, Mitsuaki and Anagnostopoulos, Konstantinos N. and Azuma, Takehiro and Hatakeyama, Kohta and Nishimura, Jun and Papadoudis, Stratos Kovalkov and Tsuchiya, Asato",
    title = "{The emergence of expanding space-time in a novel large-$N$ limit of the Lorentzian type IIB matrix model}",
    eprint = "2212.10127",
    archivePrefix = "arXiv",
    primaryClass = "hep-lat",
    reportNumber = "KEK-TH-2485",
    doi = "10.22323/1.430.0371",
    journal = "PoS",
    volume = "LATTICE2022",
    pages = "371",
    year = "2023"
}

@article{hep-th/9811220,
    author = "Hotta, Tomohiro and Nishimura, Jun and Tsuchiya, Asato",
    title = "{Dynamical aspects of large N reduced models}",
    eprint = "hep-th/9811220",
    archivePrefix = "arXiv",
    reportNumber = "UT-KOMABA-98-12, DPNU-98-22, OU-HET-298",
    doi = "10.1016/S0550-3213(99)00056-5",
    journal = "Nucl. Phys. B",
    volume = "545",
    pages = "543--575",
    year = "1999"
}

@article{hep-th/9803117,
    author = "Krauth, Werner and Nicolai, Hermann and Staudacher, Matthias",
    title = "{Monte Carlo approach to M theory}",
    eprint = "hep-th/9803117",
    archivePrefix = "arXiv",
    reportNumber = "AEI-058",
    doi = "10.1016/S0370-2693(98)00557-7",
    journal = "Phys. Lett. B",
    volume = "431",
    pages = "31--41",
    year = "1998"
}

@article{hep-th/0003223,
    author = "Nishimura, Jun and Vernizzi, Graziano",
    title = "{Spontaneous breakdown of Lorentz invariance in IIB matrix model}",
    eprint = "hep-th/0003223",
    archivePrefix = "arXiv",
    reportNumber = "NBI-HE-00-15",
    doi = "10.1088/1126-6708/2000/04/015",
    journal = "JHEP",
    volume = "04",
    pages = "015",
    year = "2000"
}

@unpublished{inprogress,
    author = "Anagnostopoulos, Konstantinos N. and Azuma, Takehiro and Hirasawa, Mitsuaki and
    Karydis, Evangelos and Nishimura, Jun and Tsuchiya, Asato and Yamamori, Naoyuki",
note = {\!\!,~work in progress}
}

@article{Bonelli:2002mb,
    author = "Bonelli, Giulio",
    title = "{Matrix strings in pp wave backgrounds from deformed super Yang-Mills theory}",
    eprint = "hep-th/0205213",
    archivePrefix = "arXiv",
    reportNumber = "ULB-TH-02-15",
    doi = "10.1088/1126-6708/2002/08/022",
    journal = "JHEP",
    volume = "08",
    pages = "022",
    year = "2002"
}

@article{Berenstein:2002jq,
    author = "Berenstein, David Eliecer and Maldacena, Juan Martin and Nastase, Horatiu Stefan",
    title = "{Strings in flat space and pp waves from N=4 super Yang-Mills}",
    eprint = "hep-th/0202021",
    archivePrefix = "arXiv",
    doi = "10.1088/1126-6708/2002/04/013",
    journal = "JHEP",
    volume = "04",
    pages = "013",
    year = "2002"
}

@article{Kumar:2022giw,
    author = "Kumar, Arpith and Joseph, Anosh and Kumar, Piyush",
    title = "{Complex Langevin study of spontaneous symmetry breaking in IKKT matrix model}",
    eprint = "2209.10494",
    archivePrefix = "arXiv",
    primaryClass = "hep-lat",
    doi = "10.22323/1.430.0213",
    journal = "PoS",
    volume = "LATTICE2022",
    pages = "213",
    year = "2023"
}

@article{Kumar:2023nya,
    author = "Kumar, Arpith and Joseph, Anosh and Kumar, Piyush",
    title = "{Investigating spontaneous SO(10) symmetry breaking in type IIB matrix model}",
    eprint = "2308.03607",
    archivePrefix = "arXiv",
    primaryClass = "hep-lat",
    doi = "10.1007/978-981-97-0289-3_337",
    journal = "Springer Proc. Phys.",
    volume = "304",
    pages = "1201--1203",
    year = "2024"
}

@article{2501_17798,
    author = "Chou, Chien-Yu and Nishimura, Jun and Tripathi, Ashutosh",
    title = "{Inequivalence between the Euclidean and Lorentzian versions of the type IIB matrix model from Lefschetz thimble calculations}",
    eprint = "2501.17798",
    archivePrefix = "arXiv",
    primaryClass = "hep-th",
    reportNumber = "KEK-TH-2686",
    doi = "10.1103/PhysRevLett.134.211601",
    journal = "Phys. Rev. Lett.",
    volume = "134",
    number = "21",
    pages = "211601",
    year = "2025"
}

@article{2407_03491,
    author = "Hirasawa, Mitsuaki and Anagnostopoulos, Konstantinos N. and Azuma, Takehiro and Hatakeyama, Kohta and Nishimura, Jun and Papadoudis, Stratos and Tsuchiya, Asato",
    title = "{The effects of SUSY on the emergent spacetime in the Lorentzian type IIB matrix model}",
    eprint = "2407.03491",
    archivePrefix = "arXiv",
    primaryClass = "hep-th",
    reportNumber = "KEK-TH-2636, KUNS-3007",
    doi = "10.22323/1.463.0257",
    journal = "PoS",
    volume = "CORFU2023",
    pages = "257",
    year = "2024"
}

@article{Iso:1999xs,
    author = "Iso, S. and Kawai, H.",
    title = "{Space-time and matter in IIB matrix model: Gauge symmetry and diffeomorphism}",
    eprint = "hep-th/9903217",
    archivePrefix = "arXiv",
    reportNumber = "KEK-TH-617",
    doi = "10.1142/S0217751X0000032X",
    journal = "Int. J. Mod. Phys. A",
    volume = "15",
    pages = "651--666",
    year = "2000"
}

@article{Ho:2025htr,
    author = "Ho, Pei-Ming and Kawai, Hikaru and Steinacker, Harold C.",
    title = "{General relativity in IIB matrix model}",
    eprint = "2509.06646",
    archivePrefix = "arXiv",
    primaryClass = "hep-th",
    reportNumber = "UWThPh 2025-17, NITEP 258",
    doi = "10.1007/JHEP02(2026)070",
    journal = "JHEP",
    volume = "02",
    pages = "070",
    year = "2026"
}

@article{Hartnoll:2025ecj,
    author = "Hartnoll, Sean A. and Liu, Jun",
    title = "{Statistical physics of the polarised IKKT matrix model}",
    eprint = "2504.06481",
    archivePrefix = "arXiv",
    primaryClass = "hep-th",
    doi = "10.21468/SciPostPhys.19.4.099",
    journal = "SciPost Phys.",
    volume = "19",
    pages = "099",
    year = "2025"
}

@article{Chou:2025rwy,
    author = "Chou, Chien-Yu and Nishimura, Jun and Wang, Cheng-Tsung",
    title = "{Monte Carlo studies of the emergent spacetime
    in the polarized type IIB matrix model}",
    eprint = "2507.18472",
    archivePrefix = "arXiv",
    primaryClass = "hep-th",
    reportNumber = "KEK-TH-2740",
    doi = "10.1103/y1rm-n85b",
    journal = "Phys. Rev. Lett.",
    volume = "135",
    number = "22",
    pages = "221601",
    year = "2025"
}

@article{Liao:2025yfb,
    author = "Liao, Henry and Maeta, Reishi",
    title = "{A new type of saddle in the Euclidean IKKT matrix model and its emergent geometry}",
    eprint = "2512.03161",
    archivePrefix = "arXiv",
    primaryClass = "hep-th",
    month = "12",
    year = "2025"
}

@article{Gass:2025bqr,
    author = "Ga{\ss}, Christian and Steinacker, Harold C.",
    title = "{Spatially flat cosmological quantum spacetimes}",
    eprint = "2510.21283",
    archivePrefix = "arXiv",
    primaryClass = "hep-th",
    doi = "10.1103/6l9d-njcr",
    journal = "Phys. Rev. D",
    volume = "113",
    number = "2",
    pages = "024050",
    year = "2026"
}

@article{Steinacker:2026qzk,
    author = "Steinacker, Harold C.",
    title = "{Modified gravity at large scales on quantum spacetime in the IKKT model}",
    eprint = "2601.08031",
    archivePrefix = "arXiv",
    primaryClass = "hep-th",
    doi = "10.1007/JHEP04(2026)044",
    journal = "JHEP",
    volume = "04",
    pages = "044",
    year = "2026"
}

@article{Hattori:2024btt,
    author = "Hattori, Keiichiro and Mizuno, Yuki and Tsuchiya, Asato",
    title = "{Regularization of matrices in the covariant derivative interpretation of matrix models}",
    eprint = "2410.13414",
    archivePrefix = "arXiv",
    primaryClass = "hep-th",
    doi = "10.1093/ptep/ptae180",
    journal = "PTEP",
    volume = "2024",
    number = "12",
    pages = "123B06",
    year = "2024"
}

@article{Gohara:2025zfh,
    author = "Gohara, Jumpei and Sako, Akifumi",
    title = "{Quantization of Lie{\textendash}Poisson algebra and Lie algebra solutions of mass-deformed type IIB matrix model}",
    eprint = "2503.24060",
    archivePrefix = "arXiv",
    primaryClass = "hep-th",
    doi = "10.1063/5.0278248",
    journal = "J. Math. Phys.",
    volume = "67",
    number = "2",
    pages = "022301",
    year = "2026"
}

@article{Laliberte:2023bai,
    author = "Laliberte, Samuel and Brahma, Suddhasattwa",
    title = "{IKKT thermodynamics and early universe cosmology}",
    eprint = "2304.10509",
    archivePrefix = "arXiv",
    primaryClass = "hep-th",
    doi = "10.1007/JHEP11(2023)161",
    journal = "JHEP",
    volume = "11",
    pages = "161",
    year = "2023"
}

@article{Mollgaard:2013qra,
    author = "Mollgaard, A. and Splittorff, K.",
    title = "{Complex Langevin dynamics for chiral Random Matrix Theory}",
    eprint = "1309.4335",
    archivePrefix = "arXiv",
    primaryClass = "hep-lat",
    doi = "10.1103/PhysRevD.88.116007",
    journal = "Phys. Rev. D",
    volume = "88",
    number = "11",
    pages = "116007",
    year = "2013"
}

@article{Ambjorn:2000bf,
    author = "Ambjorn, Jan and Anagnostopoulos, K. N. and Bietenholz, Wolfgang and Hotta, T. and Nishimura, J.",
    title = "{Large N dynamics of dimensionally reduced 4-D SU(N) superYang-Mills theory}",
    eprint = "hep-th/0003208",
    archivePrefix = "arXiv",
    reportNumber = "NBI-HE-00-14, NORDITA-2000-8-HE, UT-KOMABA-99-19",
    doi = "10.1088/1126-6708/2000/07/013",
    journal = "JHEP",
    volume = "07",
    pages = "013",
    year = "2000"
}

@article{Tsutsui:2025jez,
    author = "Tsutsui, Shoichiro and Asano, Yuhma and Ito, Yuta and Matsufuru, Hideo and Namekawa, Yusuke and Nishimura, Jun and Shimasaki, Shinji and Tsuchiya, Asato",
    title = "{On the validity of the complex Langevin method near the deconfining phase transition in QCD at finite density}",
    eprint = "2505.06551",
    archivePrefix = "arXiv",
    primaryClass = "hep-lat",
    reportNumber = "KEK-TH-2355, UTHEP-801, RIKEN-iTHEMS-Report-25",
    doi = "10.1007/JHEP10(2025)108",
    journal = "JHEP",
    volume = "10",
    pages = "108",
    year = "2025"
}

@article{Ito:2020mys,
    author = "Ito, Yuta and Matsufuru, Hideo and Namekawa, Yusuke and Nishimura, Jun and Shimasaki, Shinji and Tsuchiya, Asato and Tsutsui, Shoichiro",
    title = "{Complex Langevin calculations in QCD at finite density}",
    eprint = "2007.08778",
    archivePrefix = "arXiv",
    primaryClass = "hep-lat",
    reportNumber = "KEK-TH-2230, RIKEN-QHP-479",
    doi = "10.1007/JHEP10(2020)144",
    journal = "JHEP",
    volume = "10",
    pages = "144",
    year = "2020"
}

@article{Mandl:2025mav,
    author = "Mandl, Michael and Seiler, Erhard and Sexty, D{\'e}nes",
    title = "{Necessary and sufficient conditions for correctness of complex Langevin}",
    eprint = "2508.14512",
    archivePrefix = "arXiv",
    primaryClass = "hep-lat",
    doi = "10.1088/1751-8121/ae2245",
    journal = "J. Phys. A",
    volume = "58",
    number = "49",
    pages = "495202",
    year = "2025"
}

@article{Scherzer:2020kiu,
    author = "Scherzer, M. and Sexty, D. and Stamatescu, I. O.",
    title = "{Deconfinement transition line with the complex Langevin equation up to $\mu /T \sim 5$}",
    eprint = "2004.05372",
    archivePrefix = "arXiv",
    primaryClass = "hep-lat",
    doi = "10.1103/PhysRevD.102.014515",
    journal = "Phys. Rev. D",
    volume = "102",
    number = "1",
    pages = "014515",
    year = "2020"
}

@article{Scherzer:2019lrh,
    author = "Scherzer, M. and Seiler, E. and Sexty, D. and Stamatescu, I. -O.",
    title = "{Controlling complex Langevin simulations of lattice models by boundary term analysis}",
    eprint = "1910.09427",
    archivePrefix = "arXiv",
    primaryClass = "hep-lat",
    doi = "10.1103/PhysRevD.101.014501",
    journal = "Phys. Rev. D",
    volume = "101",
    number = "1",
    pages = "014501",
    year = "2020"
}

@article{Sexty:2019vqx,
    author = "Sexty, D{\'e}nes",
    title = "{Calculating the equation of state of dense quark-gluon plasma using the complex Langevin equation}",
    eprint = "1907.08712",
    archivePrefix = "arXiv",
    primaryClass = "hep-lat",
    doi = "10.1103/PhysRevD.100.074503",
    journal = "Phys. Rev. D",
    volume = "100",
    number = "7",
    pages = "074503",
    year = "2019"
}

@article{Aarts:2017vrv,
    author = "Aarts, Gert and Seiler, Erhard and Sexty, Denes and Stamatescu, Ion-Olimpiu",
    title = "{Complex Langevin dynamics and zeroes of the fermion determinant}",
    eprint = "1701.02322",
    archivePrefix = "arXiv",
    primaryClass = "hep-lat",
    doi = "10.1007/JHEP05(2017)044",
    journal = "JHEP",
    volume = "05",
    pages = "044",
    year = "2017",
    note = "[Erratum: JHEP 01, 128 (2018)]"
}

@article{Fodor:2015doa,
    author = {Fodor, Z. and Katz, S. D. and Sexty, D. and T{\"o}r{\"o}k, C.},
    title = "{Complex Langevin dynamics for dynamical QCD at nonzero chemical potential: A comparison with multiparameter reweighting}",
    eprint = "1508.05260",
    archivePrefix = "arXiv",
    primaryClass = "hep-lat",
    doi = "10.1103/PhysRevD.92.094516",
    journal = "Phys. Rev. D",
    volume = "92",
    number = "9",
    pages = "094516",
    year = "2015"
}

@article{Sexty:2013ica,
    author = "Sexty, D{\'e}nes",
    title = "{Simulating full QCD at nonzero density using the complex Langevin equation}",
    eprint = "1307.7748",
    archivePrefix = "arXiv",
    primaryClass = "hep-lat",
    doi = "10.1016/j.physletb.2014.01.019",
    journal = "Phys. Lett. B",
    volume = "729",
    pages = "108--111",
    year = "2014"
}

@article{Asano:2025qfb,
    author = "Asano, Yuhma and Ito, Yuta and Nishimura, Jun and Shimizu, Noritaka",
    title = "{Quantum Monte Carlo calculations in the nuclear shell model by the complex Langevin method}",
    eprint = "2508.15116",
    archivePrefix = "arXiv",
    primaryClass = "nucl-th",
    reportNumber = "UTHEP-809, KEK-TH-2747",
    month = "8",
    year = "2025"
}

@article{Mandl:2026vdc,
    author = "Mandl, Michael",
    title = "{Correctness criteria for complex Langevin}",
    eprint = "2604.12388",
    archivePrefix = "arXiv",
    primaryClass = "hep-lat",
    month = "4",
    year = "2026"
}

\end{document}